\documentclass[3p,preprint,number]{elsarticle}
\usepackage[utf8]{inputenc}
\usepackage{graphicx,xcolor}
\usepackage{amsmath,amssymb,bm} 
\usepackage{microtype}
\usepackage{caption}
\usepackage{booktabs}

\usepackage[unicode,colorlinks=true,linkcolor=blue,citecolor=blue,urlcolor=blue]{hyperref}
\usepackage[all]{hypcap} 

\newcommand{\kB}{k_{\rm B}}
\newcommand{\OmegaJ}{\Omega_{\rm J}}

\newcommand{\dd}{{\rm d}}
\renewcommand{\(}{\left(}
\renewcommand{\)}{\right)}

\begin{document}
\begin{frontmatter}

\title{Phase diffusion and noise temperature of a microwave amplifier \\ based on single unshunted Josephson junction}

%% Authors per affiliation:
\author[1]{Artem Ryabov\corref{cor1}}
\ead{artem.ryabov@mff.cuni.cz}

\author[2]{Martin \v{Z}onda}
\ead{martin.zonda@karlov.mff.cuni.cz}

\author[2]{Tom\'{a}\v{s} Novotn\'{y}}
\ead{tno@karlov.mff.cuni.cz}

%% affiliations 
\cortext[cor1]{Corresponding author}

\address[1]{Charles University, Faculty of Mathematics and Physics, Department of Macromolecular Physics, V~Hole{\v s}ovi{\v c}k{\' a}ch~2, CZ-180~00 Praha, Czech Republic}
\address[2]{Charles University, Faculty of Mathematics and Physics, Department of Condensed Matter Physics, Ke~Karlovu~5, CZ-121~16 Praha, Czech Republic}

%----------------------------------------------------------------------
%% Abstract, keywords
\begin{abstract}
High-gain microwave amplifiers operating near quantum limit are crucial for development of quantum technology. However, a systematic theoretical modeling and simulations of their performance represent rather challenging tasks due to the occurrence of colored noises and nonlinearities in the underlying circuits. 
Here, we develop a response theory for such an amplifier whose circuit dynamics is based on nonlinear oscillations of an unshunted Josephson junction. The theory accounts for a subtle interplay between exponentially damped fluctuations around the stable limit cycle and the nonlinear dynamics of the limit cycle phase. The amplifier gain and noise spectrum are derived assuming a colored voltage noise at the circuit resistor. The derived expressions are generally applicable to any system whose limit cycle dynamics is perturbed by a colored noise and a harmonic signal. We also critically assess reliabilities of numerical methods of simulations of the corresponding nonlinear Langevin equations, where even reliable discretization schemes might introduce errors significantly affecting simulated characteristics at the peak performance. \end{abstract} 

\begin{keyword} 
\texttt{Phase diffusion\sep nonlinear oscillations\sep quasi-linear response theory\sep microwave amplifiers}
\end{keyword}

\end{frontmatter}

%----------------------------------------------------------------------
\section{Introduction}

Various high-gain amplifiers operating close to the quantum limit~\cite{Caves1982, Caves2012} have been recently designed for the sake of ratcheting up weak quantum signals \cite{Bergeal2010, RevModPhys.82.1155, Nation2012, Esposito2021}. Examples include optomechanical devices~\cite{Fong2014, Ockeloen-Korppi2016}, non-linear optical~\cite{Laflamme2011, Hamerly2015, Takemura2020} and superconducting cavities~\cite{Gao2011, Wustmann2017}, kinetic inductance amplifiers~\cite{Parker2021}, superconducting quantum interference devices~\cite{Kamal2012, Sundqvist2013}, or/and superconducting circuits with Josephson junctions~\cite{Yurke1989, Kamal2009, Astafiev2010, Abdo2011, Lahteenmaki2012, Roch2012, Lahteenmaki2014, Abdo2014, Roy2018, Jebari2018,  Wustmann2019}. Such devices unavoidably operate at low temperatures, where statistics of thermal fluctuations reflect quantum effects. In this regime, fluctuations with a nonconstant spectral power naturally emerge in the experimental setup. In addition to the complex noise spectra at the device output port, the fluctuations stimulate phase diffusion in nonlinear oscillating circuits that can have a fundamental impact on the device performance. Quantitative description and control of the noisy nonlinear electronic circuits represent an important research topic of practical interest for both the basic physics research and the applied circuit engineering. 

A nearly quantum-limited amplifier based on a nonlinear dynamics of single unshunted Josephson junction has been experimentally realized and tested in works~\cite{Lahteenmaki2012} and \cite{Lahteenmaki2014}. 
A practical advantage of this setup, compared to more common parametric amplifiers based on arrays of Josephson junctions~\cite{Castellanos-Beltran/Lehnert:ApplPhysLett2007, Castellanos-Beltran/etal:NatPhys2008, Spietz/etal:ApplPhysLett2008, Teufel/etal:NatNanotechnol2009, Hatridge/etal:PRB2011} is that it does not require a microwave pumping. 
Reportedly, the circuit performance characteristics exhibit remarkable nonlinear behaviors close to the amplifier working frequency. However, a microscopic theoretical explanation for the observed effects based on the mathematical analysis of underlying nonlinear Langevin equations has been missing so far. 

In this work, we develop a response theory accounting for a colored quantum noise and phase diffusion. Utilizing this perturbation expansion, we derive analytical expressions for the amplifier gain and for its noise power spectrum. Besides giving the exact formulas, the theory provides physical insights into the nature of amplification effect and establishes its relation to properties of underlying nonlinear Langevin equations subjected to the colored noise. We verify our predictions by extensive stochastic simulations whose capability to model the process properly is also discussed. 

In Sec.~\ref{sec:amplifier}, we describe the circuit dynamics and introduce principal performance characteristics of the amplifier. Each one of the Sections~\ref{sec:quasilinear}-\ref{sec:simulations} is thematically divided into two parts: In the first, we present derivations and mathematical analysis of corresponding results while in the second, their physical significance and implications for experiments are described. Namely, in Sec.~\ref{sec:quasilinear} we present the quasi-linear response theory, derive and discuss properties of the phase diffusion coefficient and relate the current theoretical approach to previous works in the field. In Sec.~\ref{sec:gain} we analyze the amplifier gain curve. In Sec.~\ref{sec:noise} the noise power spectrum at the output is described. In Sec.~\ref{sec:simulations} possible artifacts related to the numerical integration of nonlinear Langevin equations are assessed. The work is concluded by Sec.~\ref{sec:summary} summarizing briefly the main findings and general significance of the presented methodology. 

%----------------------------------------------------------------------
\section{Amplifier characteristics and circuit dynamics}
\label{sec:amplifier}

Let us now describe main quantities of interest, explain their meaning, and formulate evolution equations governing dynamics of the amplifier circuit. The equations shall be given in dimensionless units convenient for numerical calculations. 
 
\subsection{Principal characteristics} 

A principal characteristic of the amplifier is its power gain at a given frequency, $G(\omega)$. It is given by the ratio of the output power to the input power. In a reflection amplifier, it can be estimated as 
\begin{equation}
\label{eq:G_def}
G(\omega)=|\Gamma(\omega)|^{2},
\end{equation} 
where the reflection coefficient $\Gamma(\omega)$ relates amplitudes of the input and the output voltages at a given frequency, $v_{\rm out}(\omega) = \Gamma(\omega) v_{\rm in}(\omega)$.

Amplification of weak quantum input signals is feasible provided the gain is large enough and  the noise added to the output signal $v_{\rm out}(t)$ by the amplifier itself is weak. The power spectrum of the output signal,  
\begin{equation}
\label{eq:S_vout}
S(\omega) = \int_{-\infty}^{+\infty} \dd \tau  \langle v_{\rm out}(t) v_{\rm out}(t+\tau) \rangle {\rm e}^{-j \omega \tau}, 
\end{equation}
measured at the zero input signal, $v_{\rm in}(t)=0$, is formed solely by the amplifier intrinsic noise. 
The two amplifier characteristics, $G(\omega)$ and $S(\omega)$, determine a lower bound on the energy of input signals, which can be amplified with a reasonable signal-to-noise ratio. To be visible despite the presence of the amplifier intrinsic noise, the energy of the input signal at frequency $\omega$ should be higher than  
\begin{equation}
\label{eq:Tn_def} 
\kB T_{\rm n}(\omega) = \frac{1}{Z_0} \frac{S(\omega)}{G(\omega)},
\end{equation} 
where $T_{\rm n}(\omega)$ is the so called noise temperature and $Z_0$ is the internal resistance of the input signal generator. The quantity $\kB T_{\rm n}(\omega)$ can be also understood as the energy of a harmonic input signal, which after an ideal amplification would acquire the power spectrum identical with that of the amplifier intrinsic noise. 

In addition to $G(\omega)$ and $T_{\rm n}(\omega)$, the third important quantity characterizing the amplifier performance is its bandwidth $B$. We define $B$  as the so called full width at half maximum. 
That is, $B$ is an interval of $\omega$ where the amplifier gain is not less than a half of its maximal value.

%%%%%%%%%%%%%%%%%%%%%%%%%%%%%%%%%%%%%%%%%%%%
\begin{figure}[t!]
\centering 
\includegraphics[width=0.5\linewidth]{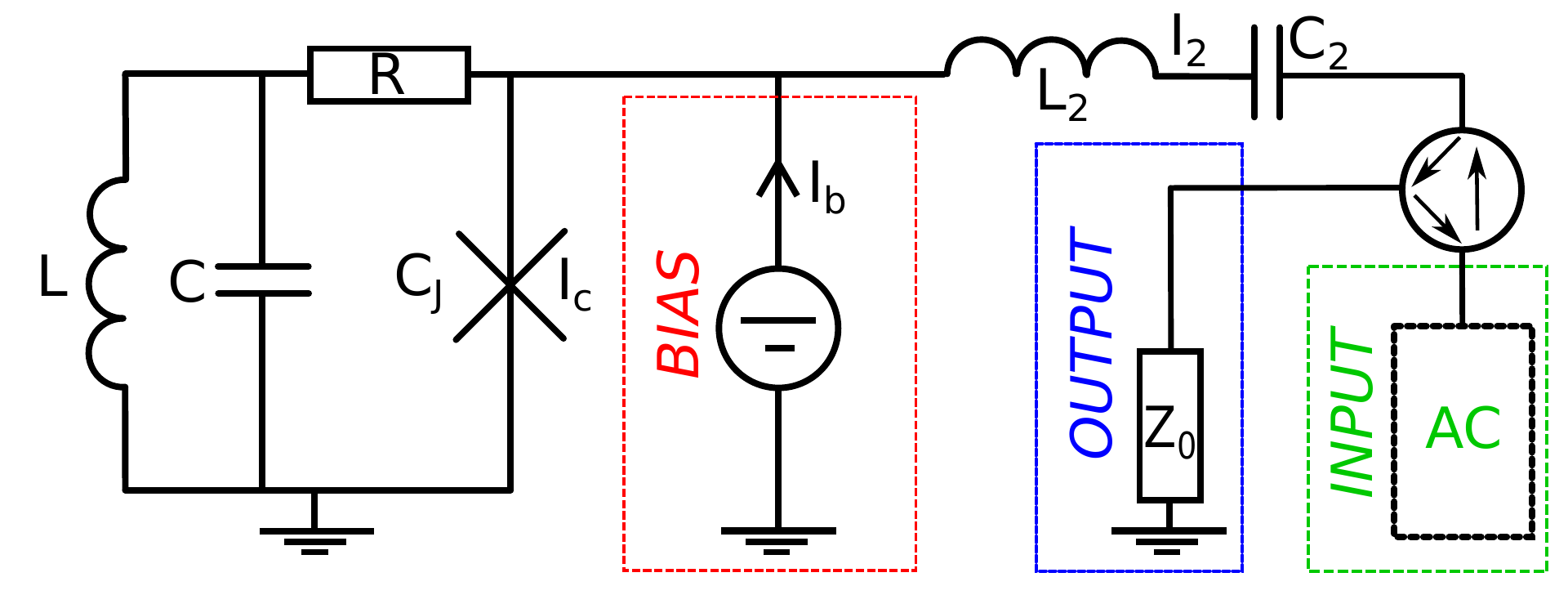}
\captionof{figure}{The simplified circuit diagram of the reflection amplifier whose detailed description can be found in the Supplementary material of Ref.~\cite{Lahteenmaki2012}. The Josephson junction is represented by a cross symbol described by $C_{\rm J}$ and $I_{\rm c}$. The resistor $R$ represents the only source of the thermal noise in the idealized model dynamics~\eqref{eq:Langevin_explicit}. Its temperature is $T\approx 4$~K, which determines the spectral power~\eqref{eq:SI}. The bias current $I_{\rm b}$ can be used to tune the operational stability and the overall amplifier performance. A harmonic input signal is generated in the right part of the circuit (AC signal generator). It then passes through the circulator (circle with arrows) and the transformer ($I_2$, $L_2$ and $C_2$). After the amplification in the left active part of the diagram, the signal is reflected back through the transformer and it is deflected by the circulator towards the output port.}
\label{fig:scheme}
\end{figure} 
%%%%%%%%%%%%%%%%%%%%%%%%%%%%%%%%%%%%%%%%%%%%

%----------------------------------------------------------------------
\subsection{Semi-classical dynamics of amplifier circuit}

Following experimental works~\cite{Lahteenmaki2012} and \cite{Lahteenmaki2014}, we model dynamics of the amplifier circuit by a system of semi-classical Langevin equations. The amplifier circuit is schematically illustrated in Fig.~\ref{fig:scheme}. Its full scheme can be found in Fig.~1 in the Supplementary material of Ref.~\cite{Lahteenmaki2012}. 
Its electronic components are described as in the classical circuit theory with the following two exceptions.  

First, the current fluctuations in the resistor $R$ are assumed to have the Callen-Welton power spectrum~\cite{Callen/Welton:1951} 
\begin{equation}
\label{eq:SI}
S_I(\omega) = \frac{2 \hslash \omega }{R} \coth\left( \frac{\hslash \omega}{2 \kB T} \right), 
\end{equation} 
accounting for both the thermal and quantum fluctuations. As a result of the microwave frequency range and the small operating temperature ($T\lesssim 4$ K), the ratio $\hslash \omega/(2\kB T)$ is not small at all relevant frequencies. Hence the classical Johnson-Nyquist spectrum~\cite{Nyquist:1928}, $S_I(\omega) = 4 \kB T/R$, which is obtained from~\eqref{eq:SI} for $\hslash \omega/(2\kB T) \ll 1$, cannot be fully justified in our case. We shall refer to the resistor current fluctuations with the spectral power~\eqref{eq:SI} as to the semi-classical resistor noise. While its spectrum originates from the quantum fluctuation-dissipation theorem, the noise process itself [denoted in the following as $\delta i(t)$] shall be described by a c-number real function of time and not by an  operator. In all numerical calculations, we set $T = 4$ K. 

Second, the unshunted Josephson junction is modeled as the circuit element whose current $I_{\rm J}(t)$ and voltage $V_{\rm J}(t)$ are determined by the Josephson phase $\varphi(t)$ via the two Josephson relations \cite{BaroneBook}  
\begin{equation}
\label{eq:IJVJ}
I_{\rm J}(t)=I_{\rm c} \sin\left( \varphi(t) \right), \qquad 
V_{\rm J}(t) = \frac{\hslash}{2e} \dot{\varphi}(t), 
\end{equation} 
respectively. Here, $e$ stands for the elementary charge and $I_{\rm c}$ is the so called critical current of the junction \cite{BaroneBook}. Furthermore, the junction has  an internal capacitance $C_{\rm J}$. 

For computational purposes, it is advantageous to formulate Langevin equations in dimensionless units. Following the experimental work~\cite{Lahteenmaki2012}, we express all currents in units of the critical current $I_{\rm c}$, voltages in units of $RI_{\rm c}$, and charges in units of $I_{\rm c}/\omega_{\rm p}$,  $\omega_{\rm p} = \sqrt{2 e I_{\rm c}/(\hbar C_{\rm J})}$. 
The plasma frequency $\omega_{\rm p}$ is also used to rescale the time variable. The Langevin equations for dimensionless quantities, formulated with the aid of Kirchhoff's laws, then read~\cite{Lahteenmaki2012} 
\begin{equation}
\label{eq:Langevin_explicit}
\frac{\dd}{\dd t} 
\left( \begin{matrix}
\varphi(t) \\ q_{\rm J}(t) \\ q(t) \\ i_L(t) \\ q_2(t) \\ i_2(t)   
\end{matrix} \right) 
= \left( \begin{matrix} 
0 & 1 & 0 & 0 & 0 & 0 \\ 
0 & -\frac{1}{Q} & \frac{C_{\rm J}}{CQ} & 0 & 0 & 1 \\
0 & \frac{1}{Q} & -\frac{C_{\rm J}}{CQ} & -1 & 0 & 0 \\
0 & 0 & \frac{\omega_0^{2}}{\omega_{\rm p}^{2}} & 0 & 0 & 0 \\ 
0 & 0 & 0 & 0 & 0 & 1 \\ 
0 & - \frac{C_2 \omega_2^{2}}{C_{\rm J} \omega_{\rm p}^{2}} & 0 & 0 & -\frac{\omega_2^{2}}{\omega_{\rm p}^{2}} & - \frac{Z_0}{L_2 \omega_{\rm p}} 
\end{matrix} \right)
\left( \begin{matrix}
\varphi(t) \\ q_{\rm J}(t) \\ q(t) \\ i_L(t) \\ q_2(t) \\ i_2(t)   
\end{matrix} \right) 
+ 
\left(\begin{matrix}
0 \\ i_{\rm b} - \delta i(t) - \sin(\varphi(t)) \\ \delta i(t) \\ 0 \\ 0 \\ \frac{2R }{ L_2 \omega_{\rm p}} v_{\rm in}(t) 
\end{matrix} \right) , 
\end{equation} 
where $i_2(t)=I_2(t) /I_{\rm c} = \dd q_2/\dd t$, $q_2(t)$ is the dimensionless charge on the capacitor $C_2$, see Fig.~\ref{fig:scheme}, $i_{L}(t)$ is the dimensionless current through the coil $L$, and $q(t)$ and $q_{\rm J}(t)$ are the dimensionless charges on the capacitors $C$ and $C_{\rm J}$, respectively.
On the right-hand side of Eq.~\eqref{eq:Langevin_explicit}, $Q$ denotes the circuit quality factor $Q\equiv \omega_{\rm p} R C_{\rm J}$, $\omega_0 \equiv 1/\sqrt{LC}$ and $\omega_2\equiv 1/\sqrt{L_2 C_2}$ are eigenfrequencies, and $Z_0$ is the input vacuum impedance. Values of circuit parameters $L$, $C$, $L_2$, $C_2$, $Z_0$, $R$, $C_{\rm J}$, and $I_{\rm c}$ will be assumed fixed and equal to those given in  Tab.~\ref{table}, which were used in the experiments~\cite{Lahteenmaki2012, Lahteenmaki2014}. On the other hand, the (dimensionless) bias current $i_{\rm b}$ is an easily tunable parameter in experiments and it will be varied when discussing the amplifier characteristics. 

Two types of weak perturbations of the nonlinear circuit dynamics are present in the second term on the right-hand side of Eq.~\eqref{eq:Langevin_explicit}: 
the resistor noise $\delta i(t)$, and the input voltage $v_{\rm in}(t)$. 
The dimensionless resistor noise $\delta i(t)$ has the power spectrum $ S_i(\omega )$ ($\omega$ is dimensionless as well) obtained after rescaling the power spectrum~\eqref{eq:SI} according to 
\begin{equation} 
S_i(\omega ) = \frac{\omega_{\rm p}}{I_{\rm c}^2} S_I(\omega \omega_{\rm p} ). 
\end{equation} 

A principal mathematical problem for the present work is to derive the power spectrum  of the current $i_2(t)$, cf.\ the 6th vector element in Eqs.~\eqref{eq:Langevin_explicit}. The current is proportional to the amplifier output voltage \cite{Lahteenmaki2012},
\begin{equation}
\label{eq:vout}
v_{\rm out}(t) =  v_{\rm in}(t) - \frac{Z_0}{R}i_{2}(t), 
\end{equation} 
and its power spectrum reflects both the autonomous limit-cycle oscillations of the amplifier circuit and the responses to the two types of weak perturbations.

%%%%%%%%%%%%%%%%%%%%%%%%%%%%%%%%%%%%%%%%%%%%
\begin{table}
		\centering
		\begin{tabular}{|lc|lc|}
			\toprule
			$Z_0$     & 50\,$\Omega$     & $R$  &  4.0\,$\Omega$\\
			$C$     & 4.26\,pF     & $C_{\rm J}$  &  0.35\,pF\\
			$C_2$     & 33\,pF     & $L$  &  702\,pH\\
			$L_2$     & 14.25\,pH     & $I_{\rm c}$  &  17\,$\mu A$\\
			$I_{\rm b}$     & 140\,$\mu$A     & $\Omega_0/2\pi $  &  2.864\,GHz\\
			$\omega_{\rm p}/2\pi$  &  61\,GHz    & $\Omega_{\rm J}/2\pi$  &  270\,GHz\\
			\bottomrule
		\end{tabular}
			\caption{Model parameters corresponding to the experimental setup of Ref.~\cite{Lahteenmaki2012}. In the experiments, the resistor $R$ was kept at the temperature $T \approx 4$~K.
			\label{table}}
\end{table}
%%%%%%%%%%%%%%%%%%%%%%%%%%%%%%%%%%%%%%%%%%%%

%----------------------------------------------------------------------
\section{Quasi-linear response to signal and noise}
\label{sec:quasilinear}

Langevin equations~\eqref{eq:Langevin_explicit} can be compactly written in the vector notation  
\begin{equation}
\label{eq:Langevin}
\dot{\boldsymbol{x}}(t)={\bm f}({\bm x}(t)) + {\bm g} v_{\rm in}(t)+ {\bm \xi}\delta i(t),
\end{equation} 
where the state vector reads 
\begin{align}
\label{eq:x}
{\bm x}(t) &= \left(\varphi(t), q_{\rm J}(t), q(t), i_L(t), q_2(t), i_2(t)\right)^{\rm T}, 
\end{align} 
and the constant vectors ${\bm g}$ and ${\bm \xi}$ respectively account for positions of the input voltage and the resistor noise as they appear in the second term on the right-hand side of~\eqref{eq:Langevin_explicit}, i.e.,   
\begin{align}
\label{eq:g}
{\bm g} &= \frac{2R}{L_2 \omega_{\rm p}} \left(0,0,0,0,0,1 \right)^{\rm T}, \\ 
\label{eq:xi} 
{\bm \xi} &= \left(0, -1, 1, 0, 0, 0\right)^{\rm T} .
\end{align} 
The vector function ${\bm f}({\bm x}(t))$ in Eq.~\eqref{eq:Langevin} is a linear combination of the elements of ${\bm x}(t)$ except for a single element that contains the Josephson nonlinearity $\sin ( \varphi(t) )$, cf.\ the second line in Eqs.~\eqref{eq:Langevin_explicit}. 

For the model parameters corresponding to the experimental setup (see Tab.~\ref{table}), a solution of the unperturbed nonlinear deterministic system $\dot{\bm x}(t)={\bm f}({\bm x}(t))$, converges towards a stable limit cycle ${\bm x}_{0}(t)$. The limit-cycle period equals to that of voltage oscillations across the Josephson junction. 
Namely, on the limit-cycle, the Josephson phase $\varphi_0(t)$ satisfies $\varphi_0(t+2\pi/\OmegaJ) = 2\pi + \varphi_0(t)$, where $\Omega_{\rm J}$ is the limit-cycle (or Josephson) frequency. Other components of the vector ${\bm x}_{0}(t)$ oscillate with the period  $2\pi/\Omega_{\rm J}$, i.e., 
${\bm x}_0(t+2\pi/\Omega_{\rm J}) = {\bm x}_0(t)+ \left(2\pi, 0, 0, 0, 0, 0\right)^{\rm T}$. 

We have performed numerical calculations of ${\bm x}_0(t)$ using the finite-difference method that utilizes three-stage Lobatto IIIa formula (bvp4c function in Matlab \cite{Kierzenka/etal:2001}). For parameters from Tab.~\ref{table}, we have obtained the Josephson frequency $\Omega_{\rm J}/2\pi \approx 270.4$~GHz, which is in agreement with its experimental value $270$ GHz given in Tab.~1 in Ref.~\cite{Lahteenmaki2012}.

In the idealized circuit model defined by Eqs.~\eqref{eq:Langevin}, the deterministic limit cycle ${\bm x}_{0}(t)$ is perturbed by the weak resistor noise $\delta i(t)$ and by the input signal $v_{\rm in}(t)$. 
The perturbations induce deviations from the limit cycle ${\bm x}_{0}(t)$ which can be of two types: 
(i) a slow diffusion of the limit cycle phase $\vartheta(t)$ and
(ii) damped amplitude fluctuations $\delta {\bm x}(t)$ characterizing deviations of ${\bm x}(t)$ from the limit cycle. The quasi-linear perturbation theory that accounts properly for both types of deviations is based on the approximation~\cite{KuramotoBook, Kaertner:1989} 
 \begin{equation} 
\label{eq:quasilin} 
{\bm x}(t) \approx {\bm x}_0(\phi(t))+\delta {\bm x}(t),
\end{equation} 
where the argument of ${\bm x}_0(.)$ accounts for the diffusing phase, 
\begin{equation}
\phi(t) = t + \vartheta(t) . 
\end{equation} 

In the Ansatz~\eqref{eq:quasilin}, small parameters are assumed to be the amplitude deviations $\delta {\bm x}(t)$ and the time-derivative of slowly diffusing phase $\dot \vartheta(t)$. 
Inserting~\eqref{eq:quasilin} into the Langevin equations~\eqref{eq:Langevin} and linearizing their right-hand sides yields  
\begin{equation} 
\label{eq:coupled_linearized}
\dot{\phi}(t) \dot{\bm x}_0(\phi(t)) + \delta \dot{\bm x}(t)  \approx 
{\bm f}\!\left( {\bm x}_0(\phi(t)) \right)+
{\bf H}(\phi(t))\cdot \delta {\bm x}(t) + {\bm g} v_{\rm in}(t)+ {\bm \xi}\delta i(t).   
\end{equation} 
The Jacobian matrix ${\bf H}(t)$ has time-periodic elements given by ${H}_{ij} = \partial {f}_{i}/\partial {x}_j$, ${\bf H}(t)={\bf H}(t+2\pi/\OmegaJ)$. The linearized expression ${\bm f}({\bm x}_0(t))+{\bf H}(t)\cdot \delta {\bm x}(t)$ contains a linear approximation of the sine nonlinearity in the Josephson phase: 
$\sin(\varphi(t)) \approx \sin(\varphi_0(t)) + \cos(\varphi_0(t)) \delta \varphi(t)$. 

%----------------------------------------------------------------------
\subsection{Decoupling of linearized equations for amplitudes and phase}

In order to decouple Eq.~\eqref{eq:coupled_linearized} into independent Langevin equations for $\vartheta(t)$ and $\delta {\bm x}(t)$, it is useful to introduce Floquet basis vectors that span solution spaces 
of the forward and backward linear differential equations
\begin{align}
\label{eq:lineqV}
& \dot{\bm v}(t) = {\bf H}(t) \cdot {\bm v}(t), \\ 
\label{eq:lineqVbar}
& \dot{\overline{\bm v}}(t) = - \overline{\bm v}(t) \cdot {\bf H}(t), 
\end{align} 
respectively. Here, the evolution operator is given by the time-periodic Jacobian matrix ${\bf H}(t)$ arising in course of the linearization~\eqref{eq:coupled_linearized}. 

The solution of system~\eqref{eq:lineqV} can be expressed as a linear combinations of 
the (column) basis vectors $\{ {\rm e}^{\lambda_n t} {\bm p}_n(t) \}_{n=0}^5$. In case of 
Eqs.~\eqref{eq:lineqVbar}, the solution can be represented by a linear combination of row vectors $\{ {\rm e}^{-\lambda_n t} \overline {\bm p}_n(t) \}_{n=0}^5$ forming the dual basis. 
Constants $\lambda_n$, known as Floquet exponents, determine stability of the solutions ${\bm v}(t)$ and $\overline {\bm v}(t)$. 
The eigenvectors ${\bm p}_n(t)$ and $\overline {\bm p}_n(t)$ are periodic with the same period as ${\bf H}(t)$, i.e., ${\bm p}_n(t+2\pi/\OmegaJ)={\bm p}_n(t)$, $\overline {\bm p}_n(t+2\pi/\OmegaJ)= \overline {\bm p}_n(t)$, and they satisfy the eigenvalue problems 
\begin{align} 
\lambda_n {\bm p}_n(t) = {\bf H}(t) \cdot {\bm p}_n(t), \\ 
\lambda_n \overline {\bm p}_n(t) = \overline {\bm p}_n(t) \cdot {\bf H}(t), 
\label{eq:pn_adjoint}
\end{align} 
for $t\in [0,2\pi/\OmegaJ]$.
In addition, we normalize the eigenvectors in such a way that the orthonormality relation 
\begin{equation}
\label{eq:normalization_pn}
\overline{\bm p}_n(t) \cdot {\bm p}_m(t) = \delta_{nm}, \quad m,n = 0,\ldots, 5,
\end{equation}
holds at all times $t$ \cite{HakenBook}. 
Such defined basis and the dual basis vectors are useful for constructing projection operators onto directions associated with the dynamics of $\vartheta(t)$ and $\delta {\bm x}(t)$. Notice that the scalar product in Eq.~\eqref{eq:normalization_pn} is defined without the complex conjugation. 

For perturbations of a stable limit cycle, it is typical that one Floquet exponent equals zero, say $\lambda_0=0$, and all other have nonzero negative real parts corresponding to the exponential decay rates of perturbations~\cite{KuramotoBook, Kaertner:1989}. 

The eigenvector to $\lambda_0=0$ is equal to the velocity vector of the unperturbed limit-cycle dynamics, i.e.,  ${\bm p}_0(t)=\dot{\bm x}_0(t)$. To see this, note that both ${\bm p}_0(t)$ and $\dot{\bm x}_0(t)$ are $(2\pi/\OmegaJ)$-periodic and that they both satisfy the linear equation~\eqref{eq:lineqV} 
[as $\dot{\bm x}_0(t)={\bm f}({\bm x}_0(t))$, we have $\ddot {\bm x}_0(t)={\bf H}(t)\cdot \dot {\bm x}_0(t)$]. Perturbations acting parallel to the limit cycle are not damped ($\lambda_0=0$) and contribute to diffusion of $\vartheta(t)$. Their magnitude is obtained by the projection 
$\overline{\bm p}_0(t) \cdot \left[ {\bm g} v_{\rm in}(t)+ {\bm \xi}\delta i(t) \right]$. 

Perturbations acting in any direction normal to the limit cycle does not contribute to the phase diffusion process $\vartheta(t)$. They result in exponentially damped amplitude deviations $\delta {\bm x}(t)$ that can be represented  as a linear combination
$\delta {\bm x}(t) = \sum_{n=1}^5 \alpha_n {\rm e}^{\lambda_n t} {\bm p}_n(\phi(t))$,
not including the undamped mode ${\bm p}_0(t)$. Therefore, the relation~\eqref{eq:normalization_pn} implies the orthogonality   
${\bm p}_0( \phi(t)) \cdot \delta {\bm x}(t)=0$, 
which we shall now use to decouple the relation~\eqref{eq:coupled_linearized} into independent Langevin equations for $\vartheta(t)$ and $\delta {\bm x}(t)$. 

Projecting both sides of Eq.~\eqref{eq:coupled_linearized} onto a direction tangent to the limit cycle, i.e., multiplying them by $\overline {\bm p}_0(\phi(t))$ from the left, and using the orthogonality 
$\overline {\bm p}_0( \phi(t)) \cdot \delta {\bm x}(t)=0$, 
gives us a closed nonlinear Langevin equation for $\vartheta(t)$. It reads 
\begin{equation} 
\label{eq:Langevin_theta}
\dot{\vartheta}(t) = Z(t+\vartheta(t)) \delta i(t),
\end{equation} 
where $Z(\phi(t))= \overline{\bm p}_0(\phi(t))\cdot {\bm \xi}$ is a magnitude of the projected resistor noise. In Eq.~\eqref{eq:Langevin_theta}, we have omitted a projection of the deterministic periodic driving, $\overline{\bm p}_0(\phi(t))\cdot {\bm g}v_{\rm in}(t)$. Its effect averages out on long time scales on which $\vartheta(t)$ evolves. Hence the driving does not affect statistics of the slow phase diffusion process $\vartheta(t)$. 

Projection of the coupled equation~\eqref{eq:coupled_linearized} onto directions perpendicular to the limit cycle gives us a closed system of Langevin equations for the amplitude deviations $\delta {\bm x}(t)$,
\begin{equation}
\label{eq:Langevin_dx}
\delta \dot{\bm x}(t) = {\bf H}(\phi(t))\cdot \delta {\bm x}(t) + {\bm \eta}(t), 
\end{equation}
where ${\bm \eta}(t)$ is the total projected perturbation onto the direction normal to the limit cycle: 
${\bm \eta}(t) = \sum_{n=1}^{5} 
{\bm p}_n(\phi(t)) \otimes \overline{\bm p}_n(\phi(t))
\cdot \left[ {\bm g}v_{\rm in}(t) + {\bm \xi}\delta i(t) \right]$. 
The symbol $\otimes$ stands for the tensor product. Its explicit evaluation is exemplified below in gain and noise calculations. 

A solution of the system of nonhomogeneous linear Langevin equations~\eqref{eq:Langevin_dx}
 acquires the form 
\begin{equation}
\label{eq:Green_qlin}
\delta {\bm x}(t) = \int_{-\infty}^{t} \dd t'\, {\bf G}(\phi(t),\phi(t')) \cdot 
\left[{\bm g}v_{\rm in}(t') + {\bm \xi}\delta i(t') \right], 
\end{equation} 
where the matrix Green function reads  
\begin{equation}
\label{eq:Green_lin}
{\bf G}(t,t') = \sum_{n=1}^{5} {\rm e}^{\lambda_n (t-t')}  {\bm p}_n(t) \otimes \overline{\bm p}_n(t'). 
\end{equation} 
The lower limit of the integration in~\eqref{eq:Green_qlin} is chosen to eliminate effects of initial conditions as we are primarily interested in a stationary response. 
In the solution~\eqref{eq:Green_qlin}, the phase diffusion process $\vartheta(t)$ entering the time variables of the Green function through the combinations $\phi(t)=t+\vartheta(t)$ is assumed to be known. Let us  now characterize $\vartheta(t)$ by deriving its diffusion coefficient. 

%----------------------------------------------------------------------
\subsection{Phase diffusion coefficient}

The limit cycle phase $\vartheta(t)$ undergoes a slow diffusion process caused by the resistor noise $\delta i (t)$. As the lowest-order approximation in the noise strength, the process $\vartheta(t)$ can be described by a Brownian motion with zero mean, 
$\langle \vartheta(t) \rangle \approx 0$, and with the variance 
\begin{align} 
\langle \vartheta^{2}(t) \rangle  \approx 2 D_\vartheta t.
\end{align} 
The phase diffusion coefficient $D_\vartheta$ depends on the spectrum of the resistor noise $\delta i (t)$. It   can be evaluated based on the expression  
\begin{equation} 
\label{eq:Dtheta_def}
D_{\vartheta} = \lim_{t\to\infty} \frac{1}{2t} 
\int_0^{t} \dd t_1 \int_0^{t} \dd t_2 \langle \dot{\vartheta}(t_1) \dot{ \vartheta}(t_2) \rangle,
\end{equation}  
into which we substitute for $\dot \vartheta(t)$ the right-hand side of the Langevin equation~\eqref{eq:Langevin_theta} and represent the periodic function $Z(t)$, $Z(\phi(t))=Z(\phi(t)+2\pi/\OmegaJ)$, by its Fourier series  
\begin{equation}
\label{eq:Zk}
Z(\phi(t)) = \sum_{k} Z[k]
\exp\!\left[j k \left( \OmegaJ  \phi(t) +   \alpha \right)\right]. 
\end{equation} 
Here, $\alpha$ is the initial phase chosen randomly in the interval $[0,2\pi]$, the summation runs over all integers $k$, and $Z[k]$ stands for the complex amplitude of the $k$th Fourier mode. 

After averaging over the random initial phase $\alpha$ and using $Z[-k]=Z^{*}[k]$, we arrive at the exact expression   
\begin{equation} 
\label{eq:thetydot}
\langle \dot{\vartheta}(t_1) \dot{\vartheta}(t_2) \rangle 
= \sum_{k} |Z[k]|^{2}
\exp\!\left[jk \OmegaJ \left( t_1 - t_2 \right)\right] 
\langle  \delta i(t_1) \delta i(t_2) 
\exp\!\left[jk \OmegaJ \left( \vartheta(t_1) - \vartheta(t_2) \right) \right]  
\rangle, 
\end{equation} 
where averages on the right-hand side involve the unknown process~$\vartheta(t)$. However, 
an approximation of these averages, valid up to the first order of the resistor noise power, may be derived by splitting the correlation on the right-hand side of Eq.~\eqref{eq:thetydot} according to 
\begin{equation} 
\label{eq:Gauss3}
\langle y_1 y_2 \exp({z y_3}) \rangle = 
( \langle y_1 y_2 \rangle + z^{2} \langle y_1 y_3 \rangle \langle y_2 y_3 \rangle ) 
\exp\!\left(\frac{z^2}{2} \langle y_3^{2} \rangle \right) .
\end{equation} 

The identity~\eqref{eq:Gauss3}  is a special case of Eq.~\eqref{eq:GaussCF} and is  
valid for zero-mean Gaussian random variables $y_i$, $i=1,2,3$. 
Comparing the average in summation in Eq.~\eqref{eq:thetydot} with that on the left-hand side of Eq.~\eqref{eq:Gauss3}, we identify 
$y_1 = \delta i(t_1)$, $y_2 = \delta i(t_2)$, $y_3 = \vartheta(t_1)-\vartheta(t_2)$, and $z=j k \OmegaJ$. Furthermore, we note that the term $\langle y_1 y_3 \rangle \langle y_2 y _3 \rangle$, i.e.,   
\begin{equation}
\label{eq:SMneglectSI2} 
\langle \delta i(t_{1}) \(\vartheta(t_1) - \vartheta(t_2)\) \rangle
\langle \delta i(t_{2}) \(\vartheta(t_1) - \vartheta(t_2)\) \rangle \sim O(S_i^2),
\end{equation}
is of the second order in the noise power and thus it is negligible compared to the two-time noise correlation function 
$\langle y_1 y_2 \rangle = \langle \delta i(t_1) \delta i(t_2) \rangle$ which is linear in $S_i$. 
Therefore, the first-order approximation for the correlation function reads 
\begin{align}
\langle \dot{\vartheta}(t_1) \dot{\vartheta}(t_2) \rangle 
 \approx \sum_{k} |Z[k]|^{2}\exp\!\left[jk \OmegaJ \left( t_1 - t_2 \right)\right] 
\langle  \delta i(t_1) \delta i(t_2) \rangle 
\exp\!\left[ -(k \OmegaJ)^2 D_\vartheta |t_1 - t_2| \right] .
\label{eq:SMsplitphase}
\end{align} 

To proceed further, we express the correlation function $\langle \delta i(t_1) \delta i (t_2) \rangle$ as the Fourier transform of $S_I(\omega)$. In the limit $t \to \infty$, only $t_1 \approx t_2$ terms  contribute significantly to the double integral in Eq.~\eqref{eq:Dtheta_def} (they are extensive in time $t$). Moreover, the Fourier coefficients $|Z[k]|$ decay exponentially with increasing $k$. Hence, we may disregard the structure factors $\exp\!\left[ -(k \OmegaJ)^2 D_\vartheta |t_1 - t_2| \right]$ as the higher-order corrections  and approximate the phase diffusion coefficient by  
\begin{equation}
\label{eq:Dtheta_Si}
D_{\vartheta} \approx  \frac{1}{2}  \sum_{k} |Z[k]|^{2}  S_i(k \Omega_{\rm J}). 
\end{equation}  

To summarize, Eq.~\eqref{eq:Dtheta_Si} represents $D_\vartheta$ as a weighted summation of the resistor noise power at integer multiples of the limit-cycle frequency $\OmegaJ$ weighted by squared amplitudes from the Fourier series representation of $Z(t)$ given in Eq.~\eqref{eq:Zk}. 
The periodic function $Z(t)= \overline{\bm p}_0(t) \cdot {\bm \xi}$ equals the scalar product of the constant vector ${\bm \xi}$ defined in Eq.~\eqref{eq:xi} and $\overline{\bm p}_0(t)$, which satisfies the adjoint eigenvalue problem~\eqref{eq:pn_adjoint} with $\lambda_0=0$. The eigenvector $\overline{\bm p}_0(t)$ is normalized according to the orthonormality relation~\eqref{eq:normalization_pn}, i.e., 
$\overline{\bm p}_0(t) \cdot \dot {\bm x}_0(t) = 1$, or, equivalently, $ \overline{\bm p}_0(t) \cdot {\bm f( {\bm x}_0(t) )} = 1$, since we have $\dot {\bm x}_0(t) = {\bm p}_0(t)$ and $\dot {\bm x}_0(t) = {\bm f}({\bm x}_0(t))$. It is enough to satisfy the normalization condition at a single time instant, e.g., $t=0$. The forward~\eqref{eq:lineqV} and backward~\eqref{eq:lineqVbar} dynamics then preserve this normalization for all $t$. 

We have calculated the amplitudes $Z[k]$ and the diffusion coefficients $D_\vartheta$ following the logic of the previous paragraph. First, the unperturbed limit cycle ${\bm x}_0(t)$ and its frequency $\OmegaJ$ were found using the Matlab bvp4c function~\cite{Kierzenka/etal:2001}. Subsequently, we have used this solution to obtain the Jacobian matrix ${\bf H}(t)$ at all times $t$, $t \in [0, 2\pi/\OmegaJ]$, and solved numerically the eigenvalue problem~\eqref{eq:pn_adjoint}. 
% To this end, we have treated Eq.~\eqref{eq:pn_adjoint} as the boundary-value problem with periodic boundary conditions $\overline{\bm p}_n(0)=\overline{\bm p}_n(2\pi/\OmegaJ)$ and again employed the bvp4c. 
For $n=0$, we have $\lambda_0 = 0$ and the eigenvalue problem~\eqref{eq:pn_adjoint} reduces to the equation 
$0 = \overline {\bm p}_0(t) \cdot {\bf H}(t)$, 
which can be understood as a boundary-value problem for the unknown row vector $\overline {\bm p}_0(t)$ subjected to periodic boundary conditions $\overline{\bm p}_0(0)=\overline{\bm p}_0(2\pi/\OmegaJ)$. We have numerically solved this boundary-value problem by employing the bvp4c function. 

The obtained eigenvector $\overline {\bm p}_0(t)$ has been normalized according to Eq.~\eqref{eq:normalization_pn} requiring $\overline{\bm p}_0(t) \cdot {\bm f}( {\bm x}_0(t)) = 1$. To compute the Fourier amplitudes $Z[k]$, we have implemented the fast-Fourier-transform-based routine with the function interpolation to the cubic order~\cite{RecipesBook}. This implementation shall be used in all computations of Fourier amplitudes needed in the following sections. 

For experimental parameters from Tab.~\ref{table} and the semi-classical resistor noise with the power spectrum~\eqref{eq:SI}, we get 
$D_\vartheta \approx 5.55 \times 10^{-5}$, 
in the dimensionless units. Calculations performed with the classical Johnson-Nyquist noise power spectrum $S_I(\omega) = 4\kB T/R$ yield $D_\vartheta \approx 5.46 \times 10^{-5}$. At zero frequency, both spectra contribute identically to $D_\vartheta$, see the $k=0$ term in Eq.~\eqref{eq:Dtheta_Si}. On the contrary, at the frequencies $\omega = \pm k \OmegaJ$ with $k>0$, the classical Johnson-Nyquist spectrum attains lower values compared to the Callen-Welton spectrum~\eqref{eq:SI}, resulting in the lower value of $D_\vartheta$ for the classical resistor noise. 

The thermal resistor noise $\delta i(t)$ represents the only noise source in the model~\eqref{eq:Langevin_explicit}. However, in an actual experiment, there might exist other relevant noise sources interfering with the amplifier dynamics. Equation~\eqref{eq:Dtheta_Si} suggests that the most pronounced contribution to $D_\vartheta$ arises from noises with a nonzero spectral power at low frequencies (the $k=0$ term in the summation). Consequently, e.g., $1/f$ electronic noises can generate an additional substantial increase of the phase diffusion coefficient $D_\vartheta$ compared to its values from the previous paragraph. 

Therefore, the calculated values of $D_\vartheta$ should be understood as ideal lower bounds on $D_\vartheta$ rather than its actual value corresponding to the discussed experimental setting. In the following, we shall treat $D_\vartheta$ as a small free parameter, and discuss behavior of the amplifier gain, bandwidth, and noise power for various values of $D_\vartheta$ chosen from the interval $D_\vartheta \in [10^{-6}, 10^{-3}]$. 

Finally, let us make a few general remarks on the computation of $Z(t)$ and $\overline{\bm p}_0(t)$. In the literature on nonlinear dynamics, the function $Z(t)$ is known as the phase response curve~\cite{WinfreeBook}. It belongs among principal characteristics of perturbed nonlinear oscillations~\citep{Johnson:1999, Brown/etal:2004, Govaerts/Sautois:2006, Nakao/etal:2010, Sacre/Sepulchre:2014}. Components of the eigenvector $\overline{\bm p}_0(t)$ describe the phase response to possible infinitesimal perturbations. 
Our computation method based on the adjoint eigenvalue problem~\eqref{eq:pn_adjoint} is closely related to the so called adjoint method of the phase response curve calculation, which is based on the numerical integration of the backward evolution equation~\eqref{eq:lineqVbar}.
Equivalently, the phase response curve $Z(t)$ can be evaluated by a so called direct method based on the measurement of $\vartheta(t)$ and $\vartheta(t+2\pi/\OmegaJ)$ for a constant perturbation $\delta i$. A thorough description and comparison of the two methods including their various numerical implementations can be found, e.g., in Refs.~\cite{Brown/etal:2004, Govaerts/Sautois:2006, Novicenko/Pyragas:2012}.

As for the circuit modeling in particular, the oscillating circuits perturbed by Gaussian white noises have been studied in Refs.~\cite{Kaertner:1989} and \cite{Demir/etal:2000}. A generalization of these approaches to the case of a colored noise was proposed in Ref.~\cite{Kaertner:1990} subjected to the assumption that the colored noise can be modeled as a solution of infinite-dimensional linear system of equations involving white noise sources. Because of this assumption, the latter theory is not applicable to our case of the quantum thermal noise. In 
further theoretical studies, phenomenological models and models aimed for specific circuits~\cite{Herzel/Razavi:1999, Magierowski/Zukotynski:2004, Agrawal/Seshia:2014} and experimental setups~\cite{Pankratz/Sanchez-Sinencio/2014, Pick/etal:2015} have been developed. These models however, do not establish a direct relation between the circuit stochastic dynamics and the noise power spectra. Our results presented below go beyond these previous works since they are valid for an arbitrary colored noise and they do not depend on the particular choice of the oscillating circuit.

%----------------------------------------------------------------------
\section{Gain, bandwidth, and instability of the limit cycle}
\label{sec:gain}
The output voltage is a superposition of the amplified input signal and the intrinsic thermal resistor noise. The noise spectral power at the amplifier output shall be discussed in Sec.~\ref{sec:noise}. 
In the present Section, we focus on a physical nature of the amplification effect and describe an influence of the phase diffusion on the amplifier gain $G(\omega)$ and bandwidth $B$.

%----------------------------------------------------------------------
\subsection{Derivation of $G(\omega)$}

The output voltage is related to the input one via Eq.~\eqref{eq:vout}. Hence a key quantity that determines $v_{\rm out}(t)$ is the current through the transformer, $i_2(t)$, see Fig.~\ref{fig:scheme}. 
The current, 
$i_2(t) = i_{20}(\phi(t)) + \delta i_2(t)$, consists of the limit cycle current oscillations $i_{20}(\phi(t))$ at frequency $\OmegaJ$ with slowly diffusing phase, and of the amplitude deviation from the limit cycle $\delta i_2(t)$. 

The function $\delta i_2(t)$ contains the full description of the amplification effect. It is included in the vector $\delta {\bm x}(t)$ as its $6$th component, cf.\ Eq.~\eqref{eq:x}. Hence from Eq.~\eqref{eq:Green_qlin} it follows that   
\begin{equation}
\label{eq:i2forG}
\delta i_2(t) = \frac{2 R}{L_2 \omega_{\rm p}} \sum_{n=1}^{5}  
\int_{-\infty}^{t} \dd t' {\rm e}^{  \lambda_n [\phi(t)-\phi(t') ]} 
p_{n6}(\phi(t)) \overline p_{n6}(\phi(t')) v_{\rm in}(t')  ,  
\end{equation} 
where scalar functions $p_{n6}(t)$ and $\overline p_{n6}(t)$ are $6$ths components of eigenvectors ${\bm p}_{n}(t)$ and $\overline{\bm p}_{n}(t)$ to the Floquet exponent $\lambda_n$, defined in 
Eqs.~\eqref{eq:lineqV} and~\eqref{eq:lineqVbar}, respectively. 

To evaluate $\langle\delta i_2(t)\rangle$, we represent the periodic function $p_{n6}(\phi(t))$ by the Fourier series 
\begin{equation}
\label{eq:SMck} 
p_{n6}(\phi(t)) = \sum_{k}  p_{n6}[k] \exp\!\left[j k \left( \OmegaJ  \phi(t) +   \alpha \right)\right],  
\end{equation} 
and similarly for $\overline p_{n6}(\phi(t))$ whose Fourier amplitudes we denote as $\overline p_{n6}[l]$. The initial phase $\alpha$, $\alpha \in [0,2\pi]$, appears in Fourier series representations of both  $p_{n6}(\phi(t))$ and  $\overline p_{n6}(\phi(t))$. 
Averaging the product $ p_{n6}(\phi(t)) \overline p_{n6}(\phi(t))$ over the initial phase implies that only summands with $l=-k$ in a series representation of $\langle\delta i(t)\rangle$ are nonzero:
\begin{equation} 
\label{eq:mean_di}
\langle \delta i_2(t) \rangle = 
\frac{2 R}{L_2 \omega_{\rm p}} \sum_{n=1}^{5} \sum_{k} p_{n6}[k] \overline p_{n6}[-k] 
\int_{-\infty}^{t}\dd t' 
\langle \exp\!\left[ ( \lambda_n + j k \OmegaJ) (\phi(t)- \phi(t'))\right] \rangle v_{\rm in}(t') . 
\end{equation} 

The mean value of the exponential in~\eqref{eq:mean_di} is evaluated using 
$\langle \exp[z(\vartheta(t) - \vartheta(t'))]  \rangle = \exp( z^2 D_\vartheta |t-t'|)$. 
Performing then the integration for the harmonic input signal 
$v_{\rm in}(t) = v_{\rm in}(\omega) e^{j\omega t}$ gives us the reflection coefficient $\Gamma(\omega)$ relating amplitudes of the input and the output voltages: $v_{\rm out}(\omega)=\Gamma(\omega) v_{\rm in}(\omega)$. 
It reads 
\begin{equation} 
\label{eq:Gamma}
 \Gamma(\omega) = 1 -  \frac{2 Z_0}{L_2 \omega_{\rm p}}
\sum_{n=1}^{5} \sum_{k}  \frac{p_{n6}[k] \overline p_{n6}[-k]}{j\omega - ( \Lambda_{nk} + D_\vartheta \Lambda_{nk}^2)}, 
\end{equation} 
where $\Lambda_{nk}$ denotes the Floquet exponent $\lambda_n$ with the imaginary part shifted by the $k$th multiple of the limit cycle frequency: 
\begin{equation}
\Lambda_{nk} = \lambda_n + jk \OmegaJ. 
\label{eq:Lambda_nk}
\end{equation}
Having obtained the reflection coefficient~\eqref{eq:Gamma}, 
the amplifier gain curve follows from Eq.~\eqref{eq:G_def} as $G(\omega) = |\Gamma(\omega)|^{2}$.

Equation~\eqref{eq:Gamma} provides important physical insights into origins and properties of the amplification effect. It relates the maximum value of the gain and the amplifier bandwidth to a measure of stability of the limit cycle. 
Moreover, it demonstrates a destructive effect of phase diffusion on the amplifier performance and an importance of sidebands [oscillations with frequencies $(\Omega_0+k\OmegaJ)$, $k=\pm 1, \pm 2,\ldots$] and various Floquet modes (characterized by exponents $\lambda_n$) for the gain values within the amplifier bandwidth. Let us now discuss these aspects. 

%----------------------------------------------------------------------
\subsection{Discussion of $G(\omega)$: Stability, amplification, relevance of sidebands and phase diffusion}
\label{sec:gain_discussion}

Real parts of Floquet exponents $\lambda_n$, $n=1,\ldots,5$, measure stability of the limit cycle as they determine damping rates of individual Floquet modes. Imaginary parts of $\lambda_n$ are equal to natural frequencies which, in general, can be different from integer multiples of the limit cycle frequency $\OmegaJ$. A resonance of the amplifier circuit occurs when an input voltage oscillates with a frequency equal to a natural frequency. If, at the same time, the corresponding damping rate is small, the resonance gives rise to a significant amplification of the input voltage amplitude. This is the principal mechanism by which the current amplifier circuit works. The phase diffusion interferes with the mechanism modifying slightly the amplifier gain and bandwidth as compared to the ordinary deterministic resonance.   

For parameter values shown in Tab.~\ref{table}, there are two Floquet exponents with a nonzero imaginary part, say $\lambda_1$ and $\lambda_2$, which are complex conjugate to each other, $\lambda_2=\lambda_1^*$. The absolute value of their imaginary parts is given by $\Omega_0$, i.e., 
\begin{equation}
 \lambda_{1,2} = -\gamma \pm j \Omega_0, 
\label{eq:lambda12}
\end{equation}
where the damping rate is $\gamma \approx 1.4 \times 10^{-5}$ (in dimensionless units) and the natural frequency $\Omega_0 \approx 4.7\times 10^{-2}$ determines value of the amplifier working frequency. The other exponents are real, negative, and they satisfy $\left|\lambda_n \right| \gg \gamma$. Namely, we have  
$\lambda_3\approx -1.8 \times 10^{-3}$, 
$\lambda_4\approx -7.6 \times 10^{-3}$, and 
$\lambda_5 \approx -2.0$. The limit cycle frequency is $\OmegaJ \approx 4.4$. 
%The Floquet modes with $\lambda_1$ and $\lambda_2$ lead to weakly damped amplitude oscillations with the working frequency $\Omega_0$ and with linear combinations $\pm \Omega_0+k\OmegaJ$. All other modes are damped more strongly and they are oscillate with integer multiples of the limit cycle frequency $\OmegaJ$. 
In SI units, the working frequency, $\Omega_0 /2\pi \approx 2.864$ GHz, and the Josephson frequency, $\Omega_{\rm J}/2\pi=270.4$ GHz, agree well with corresponding ones in experiments of Ref.~\cite{Lahteenmaki2012}, where one finds $\Omega_0 /2\pi=2865$ MHz and $\Omega_{\rm J}/2\pi=270$ GHz, see Tab.~\ref{table} in Ref.~\cite{Lahteenmaki2012}. 

A parameter that controls stability of the limit cycle and can be easily tuned in experiments is the bias current $I_{\rm b}$. If $I_{\rm b}$ is larger than a critical current $I_{\rm c}$, the voltage across the Josephson junction develops stable nonlinear oscillations, whereas at small bias currents where  $I_{\rm b}<I_{\rm c}$ the oscillations cease to exist \cite{LikharevBook}. For parameters in Tab.~\ref{table}, the critical value of the current  is approximately $I_{\rm c}\approx 0.1373$~mA. 

%%%%%%%%%%%%%%%%%%%%%%%%%%%%%%%%%%%%%%%%%%%%
\begin{figure}[t!]
\centering
\includegraphics[width=0.9\textwidth]{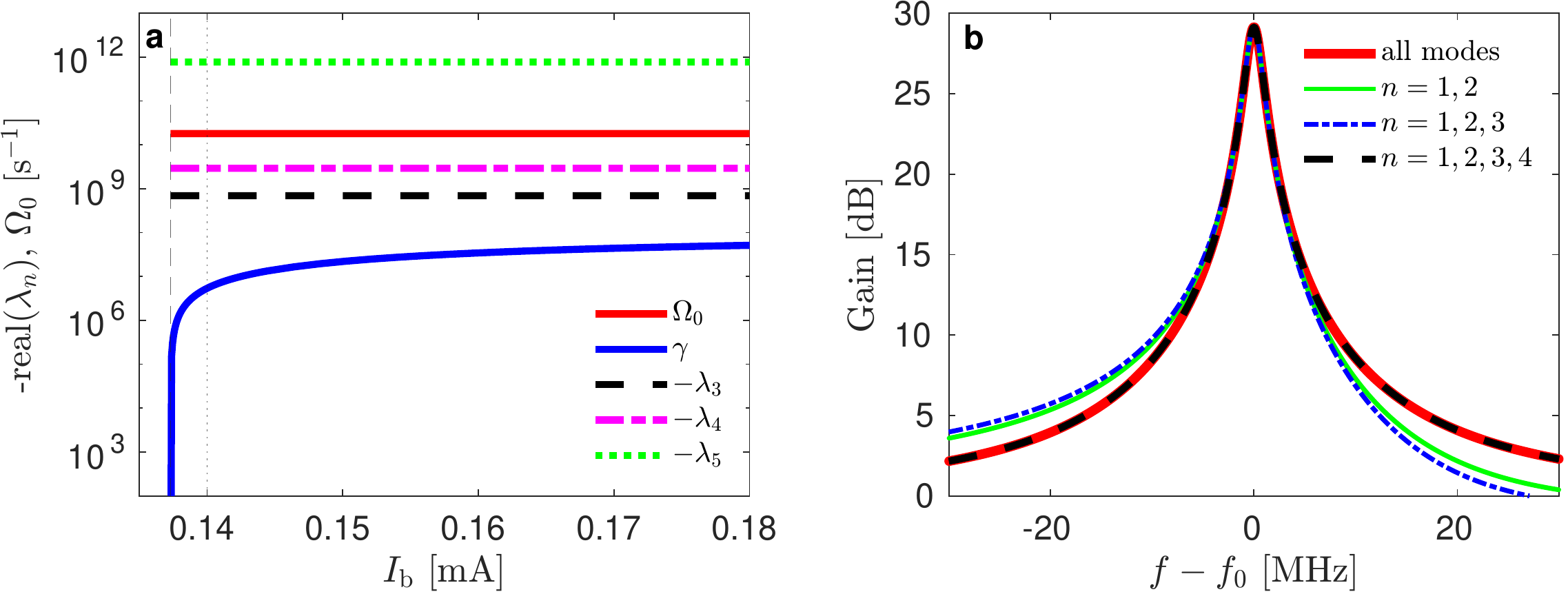} 
\caption{Panel~{\bf a}: Damping rates of the exponentially decaying Floquet modes given by the real parts of Floquet exponents $\lambda_n$ and the working frequency $\Omega_0$ of the amplifier as functions of the bias current $I_{\rm b}$. Other circuit parameters are taken from Tab.~\ref{table}. The Hopf bifurcation point, where the limit cycle becomes unstable, is denoted by the vertical thin dashed line. The vertical dotted line marks the working point $I_{\rm b}=0.14$~mA used in experiments of Ref.~\cite{Lahteenmaki2012}. Panel~{\bf b}: Consequences of discarding fast-decaying Floquet modes on the gain curve for the experimental values of circuit parameters. ``All modes'' denotes the exact gain computed based on Eq.~\eqref{eq:Gamma}. Other curves represent results obtained for various values of the mode index $n$ in Eq.~\eqref{eq:Gamma}.} 
\label{fig:gain}
\end{figure}
%%%%%%%%%%%%%%%%%%%%%%%%%%%%%%%%%%%%%%%%%%%%

Figure~\ref{fig:gain}a shows dependence of the Floquet exponents on $I_{\rm b}$. 
The critical current $I_{\rm c}$ is marked by the dashed line. Interestingly, the damping rates $-\lambda_3, -\lambda_4$, and $- \lambda_5$, and the working frequency $\Omega_0$ does not change noticeably with $I_{\rm b}$. 
Contrary, the smallest damping rate $\gamma$ vanishes when $I_{\rm b}$ approaches $I_{\rm c}$ from above and it slowly saturates with increasing $I_{\rm b}$. 
The bias used in experiments is marked by the dotted line. It is chosen close to $I_{\rm c}$ where  $\gamma$ is already small but the limit-cycle oscillations are still relatively stable in the linear regime. (For vanishingly small $\gamma$ the stability is controlled by nonlinear effects.) 

Further simplifications of the exact relation~\eqref{eq:Gamma} might explain dependencies of the amplifier gain and bandwidth on $\gamma$  and $D_\vartheta$. We shall proceed in the three steps: We neglect sidebands with nonzero $k$, show how $D_\vartheta$ gives rise to an effective damping of Floquet modes, and disregard quickly damped Floquet modes within amplifier bandwidth.  

\begin{figure}[t!]
	\centering
	\includegraphics[width=0.9\textwidth]{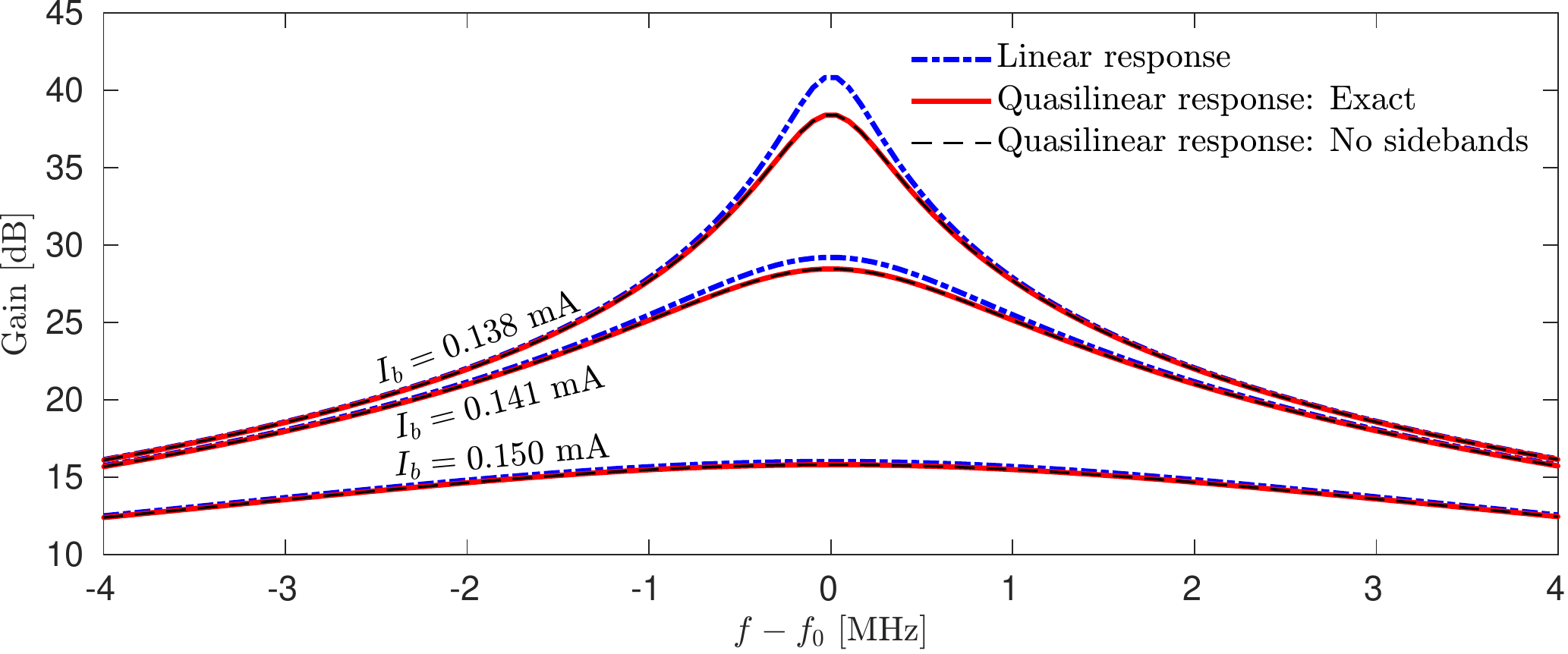} 
	\caption{Amplifier gain curves for three values of the bias current $I_{\rm b}$ controlling the limit cycle stability, viz.\ Fig.~\ref{fig:gain}{\bf a}. Solid lines represent $G(\omega)$ from Eq.~\eqref{eq:G_def} evaluated using the exact reflection coefficient~\eqref{eq:Gamma}. These exact curves overlap with the corresponding no-sideband approximations (dashed lines) obtained from the approximate reflection coefficient~\eqref{eq:Gamma_approx}. To visualize the effect of phase diffusion ($D_\vartheta=5.55\times 10^{-4}$ in dimensionless units), the linear response gain curves are plotted by dashed-dotted lines. Other parameters used are taken from Tab.~\ref{table}; $f_0=\Omega_0/2\pi$ denotes the amplifier working frequency. }
	\label{fig:gain_comparisonLR}
\end{figure}

Following this outline, we first notice that only summands with $k=0$ in Eq.~\eqref{eq:Gamma} give a significant contribution to $G(\omega)$ near $\Omega_0$. Therefore, the exact reflection coefficient~\eqref{eq:Gamma} is satisfactorily approximated by 
\begin{equation} 
\label{eq:Gamma_approx}
 \Gamma(\omega) \approx 1 -  \frac{2 Z_0}{L_2 \omega_{\rm p}}
\sum_{n=1}^{5} \frac{p_{n6}[0] \overline p_{n6}[0]}{j\omega - ( \lambda_n + D_\vartheta \lambda_n^2)}. 
\end{equation} 
The agreement between gain curves calculated from~\eqref{eq:Gamma} and~\eqref{eq:Gamma_approx} is demonstrated in Fig.~\ref{fig:gain_comparisonLR}. The two curves perfectly overlap for all tested values of $I_{\rm b}$, both close to and away from the instability. 

Second, the phase diffusion influences these both quantities increasing the damping rate by $D_\vartheta  \Omega_0^2$ for Floquet modes with exponents~\eqref{eq:lambda12}. This follows from approximations 
\begin{equation}
\label{eq:Dtheta_lambda12}
D_\vartheta \lambda_{1,2}^2 \approx - D_\vartheta \Omega_0^2, 
\end{equation}
which holds since we can neglect $ \gamma D_\vartheta$ and $ \gamma^2 D_\vartheta$ terms emerging in the products $D_\vartheta (-\gamma \pm j \Omega_0)^2$. We remind: the lower bound on $D_\vartheta$ is of the same order as $\gamma$, see the discussion after Eq.~\eqref{eq:Dtheta_Si}. Inserting into~\eqref{eq:Gamma_approx}, the term $- D_\vartheta \Omega_0^2$ adds a positive correction to the damping rate $\gamma$. Contrary, for strongly damped modes with zero natural frequencies ($n=3,4$, and 5), the phase diffusion slightly decreases the damping rates contributing by small negative corrections $D_\vartheta\lambda_n^2 > 0$.

%%%%%%%%%%%%%%%%%%%%%%%%%%%%%%%%%%%%%%%%%%%%
%%%%%%%%%%%%%%%%%%%%%%%%%%%%%%%%%%%%%%%%%%%%

Third, near the gain peak (within the amplifier bandwidth), we can further simplify the sum in~\eqref{eq:Gamma_approx} leaving out fast decaying (compared to $n=1,2$) modes with $n=3, 4$, and $5$.  
Then, near the working frequency $\Omega_0$, the gain can be approximated by the Lorentzian function  
\begin{equation}
\label{eq:G_Lorentz} 
G(\omega) \approx G(\Omega_0) 
\frac{(\gamma+D_\vartheta \Omega_0^{2} )^{2}}{(\omega-\Omega_0)^{2}
+(\gamma+D_\vartheta \Omega_0^2 )^2} ,  
\end{equation} 
where we have employed the approximation~\eqref{eq:Dtheta_lambda12}. The approximation works well within the amplifier bandwidth, while in tails it leads to an asymmetry of the gain curve with respect to the working frequency. Figure~\ref{fig:gain} shows that only the fastest mode corresponding to $\lambda_5$ does not contribute significantly to $G(\omega)$. Accounting for the modes corresponding to $\lambda_3$ and $\lambda_4$ is enough to describe the gain globally. 

The width and the height of the Lorentzian curve~\eqref{eq:G_Lorentz} are regulated by the effective damping $(\gamma+D_\vartheta \Omega_0^{2})$ since we have $ G(\Omega_0) \sim 1/(\gamma+D_\vartheta \Omega_0^{2} )^2$. We define the amplifier bandwidth $B$ as the full width at half maximum, which for Eq.~\eqref{eq:G_Lorentz} is given by $B =2( \gamma+D_\vartheta \Omega_0^{2}) $. The product of the maximum gain and the bandwidth, $B G(\Omega_0)$, stays constant with decreasing damping rate $\gamma$, i.e, as $\gamma$ tends to zero (approach to instability), the increase of the maximum gain is as pronounced as the decrease of the bandwidth. If the phase diffusion is large, it limits the maximum gain. At the same time, it helps to maintain a nonzero bandwidth even close to instability of the limit cycle.

For a comparison, a linear response theory \cite{LikharevBook, Brandt2010, Kamal2012}, which is based on the linearization of the sinus nonlinearity in Eq.~\eqref{eq:Langevin_explicit} without accounting for the phase diffusion, predicts the reflection coefficient in the form of~\eqref{eq:Gamma} with $D_\vartheta=0$. The linear-response gain has been independently verified using a standard linear response theory for circuits with Jospehson junctions~\cite{Brandt2010}. 
Close to the working frequency, the linear-response gain reads 
\begin{equation} 
\label{eq:GLR_Lorentz}
G_{\rm LR}(\omega) \approx G_{\rm LR}(\Omega_0) 
\frac{\gamma^{2}}{(\omega-\Omega_0)^{2}+\gamma^{2}}. 
\end{equation} 
As compared to Eq.~\eqref{eq:G_Lorentz}, here, the bandwidth is underestimated, therefore, $G_{\rm LR}(\omega)$ is sharper and higher than $G(\omega)$. This is demonstrated in Fig.~\ref{fig:gain_comparisonLR}, where the linear-response gain is plotted by the dashed-dotted lines. The difference between $G_{\rm LR}(\omega)$ and $G(\omega)$ is more pronounced for smaller $\gamma$, i.e., closer to the instability.

%----------------------------------------------------------------------
\section{Amplifier noise power and temperature}
\label{sec:noise}

In course of amplifier operation, the resistor noise $\delta i(t)$ is magnified and superimposed onto any output signal. To quantify the noise power at frequency $\omega$, $S(\omega)$, we evaluate the power spectrum of the output voltage $v_{\rm out}(t)$ when the input voltage is zero, $v_{\rm in}(t)=0$. In this case, Eq.~\eqref{eq:vout},  implies that the output voltage is proportional to the current $i_2(t)$,  $v_{\rm out}(t)=-(Z_0/R) i_2(t)$. 

The quasi-linear response theory~\eqref{eq:quasilin} gives us $i_2(t)$ as a superposition of the undamped oscillating current $i_{20}(\phi(t))$ with a slowly diffusing phase and of the damped oscillating amplitude $\delta i_2(t)$ obtained as the 6th component of $\delta {\bm x}(t)$ from Eq.~\eqref{eq:Green_qlin}. Accordingly, we have $v_{\rm out}(t) = v_{0}(\phi(t)) + \delta v(t)$, and the voltage $\delta v(t)$ is given by 
\begin{equation}
\label{eq:dv}
\delta v(t) =  \sum_{n=1}^{5}  
\int_{-\infty}^{t} \dd t' {\rm e}^{  \lambda_n [\phi(t)- \phi(t') ]} s_n(\phi(t)) 
\overline{s}_n(\phi(t')) \delta i(t') , 
\end{equation} 
where the periodic functions $s_n(t)$ and $\overline s_n(t)$ follow from the 6th component of the column vector ${\bm p}_{n}(t)$ multiplied by the dimensionless input impedance $(- Z_0/R)$, and from the difference of the 3rd and 2nd components of the row vector $\overline{\bm p}_{n}(t)$, respectively:  
\begin{equation}
\label{eq:SMsn}
s_n(t) = - \frac{Z_0}{R} p_{n6}(t), \quad 
\overline{s}_n(t) =  \overline p_{n3}(t) - \overline p_{n2}(t).
\end{equation} 

The two-time correlation function $\langle v_{\rm out}(t)v_{\rm out}(t+\tau) \rangle $, whose Fourier transform yields $S(\omega)$ via Eq.~\eqref{eq:S_vout}, expands into a sum of three qualitatively different correlation functions:  
\begin{equation} 
\label{eq:CF_vout}
\langle v_{\rm out}(t)v_{\rm out}(t+\tau) \rangle  = 
\left<v_0(\phi(t))v_0(\phi(t+\tau)) \right> 
+ \left< v_0(\phi(t)) \delta v(t+\tau) + v_0(\phi(t+\tau)) \delta v(t) \right> 
+ \left<\delta v(t) \delta v({t+\tau}) \right>. 
\end{equation}
Each term on the right-hand side of Eq.~\eqref{eq:CF_vout} contributes differently to $S(\omega)$: 

(i) The voltage $v_0(t)$ oscillates with the limit cycle frequency $\OmegaJ$, consequently, the Fourier transform of $\left<v_0(\phi(t))v_0(\phi(t+\tau)) \right>$ consists of a series of spectral lines of a narrow Lorentzian shape. The Lorentzian functions are centered at frequencies $k\OmegaJ$, $k = \pm 1, \pm 2, \ldots$ Their widths are determined by the phase diffusion coefficient $D_\vartheta$, see Refs.~\cite{Lax:1967, Ham/Hajimiri:2003, Levinson:2003} for a thorough discussion. 

(ii) The correlation function 
$ \left< v_0(\phi(t)) \delta v(t+\tau) + v_0(\phi(t+\tau)) \delta v(t) \right>$ is of the second order in the resistor noise power, hence its contribution to $S(\omega)$ is negligible. This fact can be shown by analogous simplifications of the correlations of Gaussian random variables as we have used to replace the correlation function in Eq.~\eqref{eq:thetydot} by its first-order approximation in Eq.~\eqref{eq:SMsplitphase}. 

(iii) The voltage $\delta v(t)$, Eq.~\eqref{eq:dv}, results from a linear combination of damped modes with frequencies $k\OmegaJ$, and of slow (weakly damped) modes with the working frequency $\Omega_0$ and combinations $(\Omega_0+k\OmegaJ)$. Therefore, within the amplifier bandwidth, the only significant contribution to the power spectrum $S(\omega)$ arises from the Fourier transform of  
$\left<\delta v(t) \delta v({t+\tau}) \right>$, and we can replace the general definition of $S(\omega)$ in Eq.~\eqref{eq:S_vout} by  
\begin{equation}
\label{eq:S_dv} 
S(\omega) = \int_{-\infty }^{+\infty} \dd \tau \langle \delta v(t) \delta v(t+\tau) \rangle 
{\rm e}^{- j \omega \tau} . 
\end{equation}

%----------------------------------------------------------------------
\subsection{Derivation of $S(\omega)$}

To evaluate the autocorrelation function in Eq.~\eqref{eq:S_dv}, we represent periodic functions $s_n(t)$ and $\overline s_n(t)$ in Eq.~\eqref{eq:dv} by Fourier series with complex amplitudes $s_n[k]$ and $\overline s_n[l]$, respectively. The Fourier series are defined like the one in Eq.~\eqref{eq:SMck} including the initial phase $\alpha$. After averaging over $\alpha$, we get  
\begin{align} 
\begin{split} 
\left< \delta v(t) \delta v(t+\tau) \right> & = \!
\sum_{n_i=1}^{5} \sum_{k_i, l_i} s_{n_1}[k_1] 
\overline{s}_{n_1}[l_1] s_{n_2}[k_2] \overline{s}_{n_2}[l_2] \delta_{k_1+k_2+l_1+l_2,0} 
\int_{-\infty}^{t+\tau}\dd t_1 \int_{-\infty}^{t} \dd t_2  
\left\langle 
\delta i(t_1) \delta i(t_2) \right. \\ 
\times & \left. 
{\rm exp}\!\left\{ 
\Lambda_{n_1 k_1} [ \phi(t+\tau) - \phi(t_1)] + \Lambda_{n_2 k_2} [ \phi(t) - \phi(t_2) ] + j (k_1+l_1) \OmegaJ [\phi(t_1) - \phi(t_2)] 
\right\} \right\rangle . 
\end{split} 
\label{eq:CF_dvdv}
\end{align} 
Here, we have used the shorthand $\Lambda_{n k}=\lambda_n + jk \OmegaJ$ from Eq.~\eqref{eq:Lambda_nk} denoting the Floquet exponents with shifted imaginary parts. 
The summations are performed over $n_1$, $n_2$ numbering Floquet modes, and integers $k_1$, $k_2$, $l_1$, and $l_2$. As a consequence of the averaging over $\alpha$, summands with $k_1+k_2+l_1+l_2 \neq 0$ are all equal to zero.   

To proceed further, we simplify the correlation functions in Eq.~\eqref{eq:CF_dvdv} with the help of the identity 
\begin{equation}  
\left< y_{1} y_{2} \exp\!\left(  \sum_{n>2} z_n y_n \right)\! \right>
= \left( \left<y_{1}y_{2} \right> +
\sum_{n,n'>2} \langle y_1 y_n \rangle \langle y_2 y_{n'} \rangle z_n z_{n'} 
\right) 
\exp\!\left(  \frac{1}{2}\sum_{k,k'>2} z_{k} \langle y_k y_{k'} \rangle z_{k'}\right),
\label{eq:GaussCF}
\end{equation} 
valid for zero-mean Gaussian random variables $y_i$ and complex numbers $z_i$, $i=1,\ldots, 5$. In our case, we identify 
$y_1=\delta i(t_1)$, 
$y_2=\delta i(t_2)$, 
$y_3=\vartheta(t+\tau) - \vartheta(t_1)$, 
$y_4=\vartheta(t) - \vartheta(t_2)$, 
$y_5=\vartheta(t_1) - \vartheta(t_2)$, 
and the corresponding complex numbers are 
$z_3= \Lambda_{n_1 k_1}$, 
$z_4= \Lambda_{n_2 k_2}$, and 
$z_5= j(k_1+l_1)\OmegaJ$. 
The sum in the exponential contains variances 
$\langle y_3^2 \rangle = 2 D_\vartheta (t+\tau-t_1)$,
$\langle y_4^2 \rangle = 2 D_\vartheta (t-t_2) $, and 
$\langle y_5^2 \rangle = 2 D_\vartheta |t_1-t_2|$, 
the correlation function 
$\langle y_4 y_5 \rangle = \theta(t_1-t_2) 2D_\vartheta  (t_1-t_2)$, 
which is identical for both $\tau>0$, $\tau<0$, 
and the correlation functions  
$\langle y_3 y_4 \rangle = \theta(\tau)\theta(t-t_1) 2D_\vartheta(t-{\rm max}\{t_1,t_2 \}) + \theta(-\tau)\theta(t+\tau-t_2)2 D_\vartheta(t+\tau-{\rm max}\{t_1,t_2 \})$, and 
$\langle y_3 y_5 \rangle = \theta(\tau) \theta(t_2-t_1) 2D_\vartheta (t_1-t_2) + \theta(-\tau)\theta(t_2-t_1) 2D_\vartheta (t_1 - {\rm min}\{t+\tau,t_2\})$.  

The preexponential factor $\langle y_1 y_2 \rangle$ on the right-hand side of Eq.~\eqref{eq:GaussCF} is proportional to the resistor noise power spectrum $S_i(\omega)$. 
The other terms, arising from the product 
$\sum_{n,n'>2} \langle y_1 y_n \rangle \langle y_2 y_{n'} \rangle z_n z_{n'}$, are of the second order in the noise power as each of them is formed by a product of two first-order terms. 
Therefore, we can approximate the correlation function in Eq.~\eqref{eq:GaussCF} by the expression 
$ \left<y_{1}y_{2} \right> 
\exp\!\left(  \sum_{k,k'>2} z_{k} \langle y_k y_{k'} \rangle z_{k'} /2\right)$. 

Employing these simplifications and writing formally the correlation function $\left<\delta i(t_1)\delta i(t_2) \right> $ as the Fourier transform of the resistor noise power spectrum, we may evaluate the double integral in Eq.~\eqref{eq:CF_dvdv}. For $\tau > 0$, the integrals in question read  
\begin{align}
\begin{split}
C_+(\tau)=&
\int_{-\infty}^{+\infty} \frac{\dd \Omega }{2\pi} S_i(\Omega) 
\int_{-\infty}^{t+\tau}\dd t_1 \int_{-\infty}^{t} \dd t_2   
\exp[ \Gamma_1(t+\tau-t_1) + \Gamma_2(t-t_2) - \varepsilon |t_1-t_2| ] \\
& \times \exp\!\left\{j [ \Omega + (k_1+l_1)\OmegaJ 
(1+ \theta(t_2-t_1)2 D_\vartheta\Lambda_{n_1k_1}+ \theta(t_1-t_2)2 D_\vartheta \Lambda_{n_2k_2})] (t_1-t_2)  \right\} \\
& \times \exp\!\left[2 \theta(t-t_1) D_\vartheta  \Lambda_{n_1k_1}\Lambda_{n_2k_2} (t-{\rm max}\{t_1 , t_2\}) \right],
\end{split}
\label{eq:C_plus}
\end{align} 
where  
$\Gamma_i = \Lambda_{n_ik_i}(1+ D_\vartheta \Lambda_{n_ik_i})$ and  
$\varepsilon =   D_\vartheta (k_1+l_1)^2\OmegaJ^2$. 

A somewhat cumbersome exponent in Eq.~\eqref{eq:C_plus} simplifies if we split the double integral into two parts:  
$\int_{-\infty}^{t+\tau}\dd t_1 \int_{-\infty}^{t} \dd t_2 = 
\int_{t}^{t+\tau} \dd t_1 \int_{-\infty}^{t} \dd t_2   
+\int_{-\infty}^{t}\dd t_1 \int_{-\infty}^{t} \dd t_2 $, and, subsequently,  
$\int_{-\infty}^{t}\dd t_1 \int_{-\infty}^{t} \dd t_2
= \int_{-\infty}^{t}\dd t_1 \int_{-\infty}^{t_1} \dd t_2 
+\int_{-\infty}^{t}\dd t_1 \int_{t_1}^{t} \dd t_2$. 
Direct evaluations of these integrals give us the final result expressed as a Fourier transform of $S_i(\Omega)$ modulated by Lorentzian-like functions: 
\begin{equation}
C_+(\tau) = 
\int_{-\infty}^{+\infty} \frac{\dd \Omega }{2\pi} S_i(\Omega) 
\frac{\exp[-\varepsilon \tau + j ( \Omega + \tilde \Omega )\tau ]}
{[\Gamma_1 +\varepsilon - j ( \Omega + \tilde \Omega ) ][\Gamma_2 -\varepsilon + j ( \Omega + \tilde \Omega ) ]}
+ O(D_\vartheta S_i), 
\qquad \tau >0,
\label{eq:C_plus_final}
\end{equation}
with $\tilde \Omega = (k_1+l_1)\OmegaJ(1+2D_\vartheta \Lambda_{n_2k_2})$. 
The symbol $O(D_\vartheta S_i)$ groups all terms containing the product $D_\vartheta S_i(\Omega)$ that are of the second order in the noise power.\footnote{To get a better insight into the structure of Eq.~\eqref{eq:C_plus_final}, one can 
perform the time integrals in~\eqref{eq:C_plus} assuming that $D_\vartheta = 0$ in all terms except for $\Gamma_i$. Straightforward calculation then yields the expression 
\begin{equation*} 
\int_{-\infty}^{+\infty} \frac{\dd \Omega }{2\pi} S_i(\Omega) 
\left\{
\frac{\exp[-\varepsilon \tau + j ( \Omega + \tilde \Omega )\tau ]}
{[\Gamma_1 +\varepsilon - j ( \Omega + \tilde \Omega ) ][\Gamma_2 -\varepsilon + j ( \Omega + \tilde \Omega ) ]}
+ \frac{2 \varepsilon \exp( \Gamma_1 \tau)}{(\Gamma_1+\Gamma_2)[\Gamma_1 - \varepsilon - j ( \Omega + \tilde \Omega )][\Gamma_2 + \varepsilon - j ( \Omega + \tilde \Omega )]} 
\right\},
\end{equation*}
that displays explicitly the $O(D_\vartheta S_i)$ term. Compared to this heuristic calculation, the nonzero $D_\vartheta$ in the exact derivation causes small shifts in dampings and resonant frequencies thus increasing an algebraic complexity of the final result. Except for these shifts, however, the nonzero $D_\vartheta$ does not introduce fundamentally new terms of the order $O(S_i)$.} 

A remaining step in derivation of the amplifier noise spectrum~\eqref{eq:S_dv} involves the integration  
\begin{equation}
F_{+}(\omega) = \int_{0}^{+\infty}\dd \tau\, C_+(\tau) {\rm e}^{-j \omega \tau}. 
\end{equation} 
Interchanging integrals over $\tau$ and $\Omega$ and performing the $\tau$ integration, we arrive at  
\begin{equation}
F_{+}(\omega) = \frac{j }{2\pi } 
\int_{-\infty}^{+\infty}\! \dd \Omega\, \frac{g(\Omega)}{j\varepsilon + \Omega - (\omega - \tilde \Omega)}
=
\frac{1}{2} g(\omega-\tilde \Omega-j\varepsilon) + \frac{j}{2 \pi} {\rm VP}\!\! \int_{-\infty}^{+\infty} \dd x\, \frac{g(x +  \omega-\tilde \Omega-j\varepsilon)}{x}. 
\label{eq:Sokhotki}
\end{equation} 
The function 
$g(\Omega)= S_i(\Omega)/\{[\Gamma_1+\varepsilon-j(\Omega+\tilde\Omega)][\Gamma_2-\varepsilon+j(\Omega+\tilde \Omega)]\}$ is a decreasing function of $|\Omega|$ and it is analytic at $\Omega = \omega-\tilde \Omega-j\varepsilon $, where the integrand has a simple pole. The imaginary part of the pole position, $-{\rm Im}[\tilde \Omega+j\varepsilon]$, is of the order $O(D_\vartheta)$. 
Provided  the pole is the integrand singularity that is closest to the real line, we may shift the integration contour from the line ${\rm Im}[\Omega]=0$ to  ${\rm Im}[\Omega]= - {\rm Im}[\tilde \Omega+j\varepsilon]$ encircling the pole from above. The integral over such deformed contour yields the two terms  on the right-hand side of Eq.~\eqref{eq:Sokhotki}, where VP denotes the Cauchy principal value. Notice that in the limit $\varepsilon \to 0$, Eq.~\eqref{eq:Sokhotki} is equivalent to the Sokhotski–Plemelj theorem.    

Computations for $\tau < 0$ run along similar lines. The corresponding $F_{-}(\omega)$ is a complex conjugate of 
$F_{+}(\omega)$. Summing the two parts,  we obtain the final result 
\begin{align} 
\begin{split} 
S(\omega )  = & 
\sum_{n_1,n_2=1}^{5} \sum_{k_1 k_2} \sum_{l_1 l_2} s_{n_1}[k_1] 
\overline{s}_{n_1}[l_1] s_{n_2}[k_2] \overline{s}_{n_2}[l_2] \delta_{k_1+k_2+l_1+l_2,0} 
\\ & \quad \times 
 \frac{{\rm Re}[ S_i(\omega - (k_1+l_1)\OmegaJ (1+ 2 \Lambda_{n_2k_2} D_{\vartheta} ) - j D_{\vartheta} (k_1+l_1)^2\OmegaJ^2 ) ]}
{[\Lambda_{n_1k_1}(1+ D_{\vartheta} \Lambda_{n_1k_1}) -j \omega]
[\Lambda_{n_2k_2}(1+ D_{\vartheta} \Lambda_{n_2k_2}) +j \omega]}, 
\end{split} 
\label{eq:S_qlin}
\end{align} 
where we have used $\Lambda_{nk} = \lambda_n + jk \OmegaJ$ as defined in Eq.~\eqref{eq:Lambda_nk}.

%----------------------------------------------------------------------
\subsection{Discussion of $S(\omega )$: Quantum effects, noise  temperature and noise suppression} 

The exact expression~\eqref{eq:S_qlin} is valid for all $\omega$. However, at frequencies within the amplifier bandwidth, only a few terms of the sum~\eqref{eq:S_qlin} yield significant contributions to the total noise power. 
Particularly, without loss of precision, we can neglect all terms with $k_1 \neq 0$ and $k_2 \neq 0$ as they are vanishingly small near~$\Omega_0$. Their smallness stems from
 the decrease of Fourier amplitudes with an increasing mode number $k$  (compared to the amplitudes with $k=0$) and from a smaller magnitude of the Lorentzian functions centered at $(\Omega_0 \pm k \OmegaJ)$ compared to the ones centered at $\Omega_0$. 
 
Setting $k_1=k_2=0$ removes also one of infinite sums over $l$, since the Kronecker delta in~\eqref{eq:S_qlin} then eliminates all terms except for the ones with $l_2 = -l_1$. Moreover, only a few values of $l_1$ are important because of the fast decrease of corresponding Fourier amplitudes with increasing $l_1$. In practice, it is sufficient to consider the terms  $l_1=0$, and $l_1=\pm 1$ only, accounting thus for the resistor noise at the working frequency, $S_i(\Omega_0)$, and at frequencies of the two sidebands, $S_i(\Omega_0\pm\OmegaJ)$ (plus small frequency shifts caused by $D_\vartheta$). 
Furthermore, we can neglect all Floquet modes except for the slowest two, which decay with exponents $\lambda_1$ and $\lambda_2$ given in Eq.~\eqref{eq:lambda12}. 

After these approximations, the infinite summations in~\eqref{eq:S_qlin} reduce to 
\begin{align} 
\begin{split} 
S(\omega )  \approx 
\sum_{n_1,n_2=1}^{2}  s_{n_1}[0]s_{n_2}[0] \sum_{l_1=-1}^{1} \overline s_{n_1}[l_1]  \overline{s}_{n_2}[-l_1] 
 \frac{{\rm Re}[ S_i(\omega - l_1 \OmegaJ -  D_\vartheta l_1 \OmegaJ ( 2 \lambda_{n_2} + j l_1 \OmegaJ ) ]} 
{[\lambda_{n_1}(1+ D_{\vartheta} \lambda_{n_1}) -j \omega]
[\lambda_{n_2}(1+ D_{\vartheta} \lambda_{n_2}) +j \omega]}. 
\end{split} 
\label{eq:S_qlin_simplified}
\end{align} 
We have tested validity of this approximation comparing it to the exact noise spectrum~\eqref{eq:S_qlin} for values of $I_{\rm b}$ used in Fig.~\ref{fig:gain_comparisonLR}. Other parameters were chosen from Tab.~\ref{table}.  
No significant deviations between the two curves occur over the $30$ MHz range around the working frequency. This range exceeds intervals of frequencies used in all figures and is significantly larger than the bandwidth. 

Therefore, Eq.~\eqref{eq:S_qlin_simplified} provides a complete description of the amplifier noise power spectrum at frequencies $\omega$ falling within the amplifier bandwidth. 
For the sake of clarity of its physical meaning, it is helpful to compare $S(\omega)$ with $S_{\rm LR}(\omega)$ computed using the linear response theory not accounting for the phase diffusion. The spectrum $S_{\rm LR}(\omega)$ is formally obtained from Eq.~\eqref{eq:S_qlin_simplified} after taking $D_\vartheta=0$. This yields    
\begin{align} 
\begin{split} 
S_{\rm LR}(\omega )  \approx 
\sum_{n_1,n_2=1}^{2}  s_{n_1}[0]s_{n_2}[0] \sum_{l_1=-1}^{1} \overline s_{n_1}[l_1]  \overline{s}_{n_2}[-l_1] 
 \frac{S_i(\omega - l_1 \OmegaJ)}
{(\lambda_{n_1} - j \omega) 
(\lambda_{n_2} + j \omega)}.
\end{split} 
\label{eq:S_lin_simplified}
\end{align} 

%%%%%%%%%%%%%%%%%%%%%%%%%%%%%%%%%%%%%%%%%%%%
\begin{figure}[t!]
\centering
\includegraphics[width=0.9\textwidth]{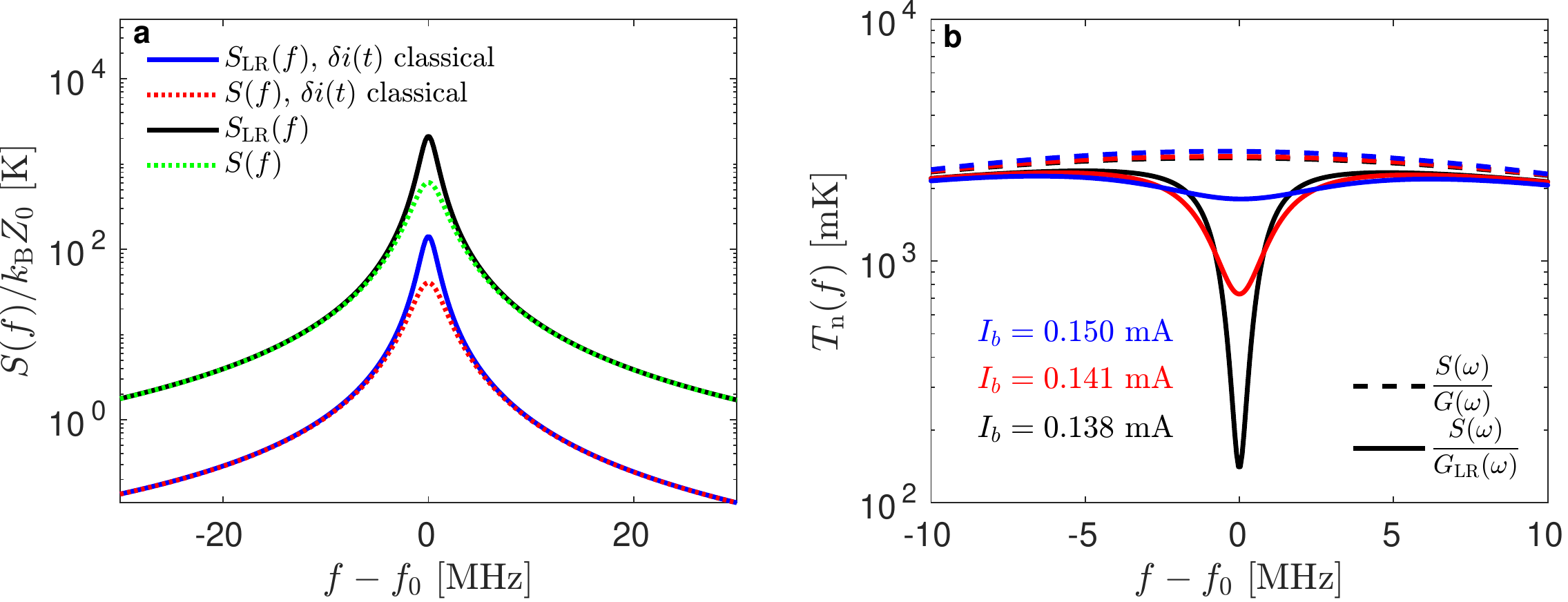} 
\caption{Panel~{\bf a}: Spectral power of the noise at the amplifier output computed from Eqs.~\eqref{eq:S_qlin_simplified} and \eqref{eq:S_lin_simplified}. The attribute ``$\delta i(t)$ classical'' (two lower curves) denotes two noise powers arising from the classical resistor current fluctuations characterized by the white Johnson-Nyquist power spectrum. In calculations of the two other curves, the Callen-Welton spectrum~\eqref{eq:SI} has been used. In this panel, all circuit parameters are taken from Tab.~\ref{table} and $D_\vartheta = 5.55 \times 10^{-3}$ in dimensionless units.  
The panel~{\bf b} demonstrates noise temperatures defined in Eq.~\eqref{eq:Tn_def}  (dashed lines) for three values of the bias current $I_{\rm b}$. Other parameters are the same as in the panel~{\bf a}. For a comparison, the corresponding three modified noise temperatures, defined with the aid of the linear response gain instead of the quasi-linear response one, are shown by three solid lines. These three curves exhibit a minimum close to the amplifier working frequency $f_0=\Omega_0/2\pi$. The most pronounced minimum is observed closest to the instability for $I_{\rm b}=0.138$~mA. The shallowest minimum occurs for $I_{\rm b}=0.150$~mA.} 
\label{fig:noise}
\end{figure}
%%%%%%%%%%%%%%%%%%%%%%%%%%%%%%%%%%%%%%%%%%%%

The noise spectrum~\eqref{eq:S_qlin_simplified} and its linear approximation~\eqref{eq:S_lin_simplified} are plotted in Fig.~\ref{fig:noise}{\bf a} for both the semi-classical resistor noise with the frequency-dependent Callen-Welton spectrum~\eqref{eq:SI} and for the classical one with the Johnson-Nyquist spectrum 
$S_I(\omega ) = 4 \kB T/R$. 
The semi-classical noise results in an output power by an order of magnitude larger compared to the classical one. 
The difference stems from the power of quantum fluctuations present in the Callen-Welton spectrum~\eqref{eq:SI}. In particular, it originates from sidebands at frequencies $\omega \approx (\Omega_0 \pm \OmegaJ)$ where the ratio 
$\hslash \omega/ (2\kB T)$ appearing as an argument of coth function in~\eqref{eq:SI} is not small.  
Indeed, at the working frequency, we have 
$\hslash \Omega_0/(2\kB T) \approx 0.17$,
and the semi-classical resistor noise power is comparable to the classical one 
(the former is higher by 1\% only). 
On the other hand, at the sideband frequency $(\Omega_0+\OmegaJ)$, we get 
$\hslash (\Omega_0+\OmegaJ)/(2 \kB T) \approx 16.39$, 
implying the semi-classical noise power that is more than an order of magnitude higher than the power of the  classical noise. 

This difference in $S(\omega)$ emphasizes the important role played by quantum fluctuations in the amplifier dynamics and performance. Qualitatively similar down-conversion of the quantum noise has been reported in experiments on current-biased Josephson junctions, see e.g.\ Ref.~\cite{Koch/etal:PRB1982}. 

The noise temperature $T_{\rm n}(\omega) = S(\omega )/[G(\omega)Z_0 \kB]$, cf.\ Eq.~\eqref{eq:Tn_def}, calculated based on the Callen-Welton spectrum~\eqref{eq:SI} varies between 1.1~K and 2.7 K within the amplifier bandwidth as demonstrated by black dashed lines in Fig.~\ref{fig:noise}{\bf b}. This is in a reasonable agreement with $T_{\rm n} \approx 2.4$ K reported in experiments~\cite{Lahteenmaki2012, Lahteenmaki2014}. Contrary, the Johnson-Nyquist spectrum results in a significantly lower noise temperature about 0.1~K (data not shown). The quantum limit of the noise temperature is $\hbar\Omega_0/2k_{\rm B} \approx 68.7$~mK.

Another remarkable feature shown in Fig.~\ref{fig:noise}{\bf b} is that the noise temperature curve is affected only weakly by the stability of the limit cycle (controlled by $I_{\rm b}$; see dashed lines). 
This constancy of $T_{\rm n}(\omega)$ is in parallel with the behavior of the gain-bandwidth product, which likewise stays nearly constant with respect to changes of~$I_{\rm b}$. 
Mathematically, this constancy can be justified after we approximate $S(\omega)$ by the Lorentzian function 
\begin{equation} 
S(\omega) \approx S(\Omega_0) \frac{(\gamma+D_\vartheta \Omega_0^{2} )^{2}}{(\omega-\Omega_0)^{2}+(\gamma+D_\vartheta \Omega_0^{2} )^{2}}, 
\label{eq:S_Lorentzian}
\end{equation} 
which is valid for $\omega$ near $\Omega_0$. 
Here, the phase diffusion yields an additional effective damping 
$D_{\vartheta} \Omega_{0}^{2}$, similarly as in the case of gain curve~\eqref{eq:G_Lorentz}. The noise temperature computed from Eqs.~\eqref{eq:S_Lorentzian} and~\eqref{eq:G_Lorentz} gives us the constant $T_{\rm n}(\omega) = S(\Omega_0)/[G(\Omega_0) Z_0 \kB]$, because the Lorentzian function in~\eqref{eq:S_Lorentzian} is identical to the one in the Lorentzian approximation of the gain in Eq.~\eqref{eq:G_Lorentz}. 

The increased effective damping rate $(\gamma + D_\vartheta \Omega_0^2)$ suppresses the noise power and broadens $S(\omega )$ for $\omega$ near $\Omega_0$ as compared to the linear-response result $S_{\rm LR}(\omega)$, which can be obtained from $S(\omega )$ by setting $D_\vartheta=0$. The broadening would occur for any type of the resistor noise, hence in Fig.~\ref{fig:noise}{\bf a} it can be seen for both the classical and the quantum case. Alternatively, the broadening manifests itself as a minimum in the noise temperature defined as $S(\omega)/[G_{\rm LR}(\omega) Z_0 \kB]$ and illustrated in Fig.~\ref{fig:noise}{\bf b}. The minimum stems from the division of the broader Lorentzian-shaped noise spectral curve~\eqref{eq:S_Lorentzian} by the narrower gain curve from Eq.~\eqref{eq:GLR_Lorentz}. It becomes more pronounced the closer to instability the limit cycle is as controlled by the bias current in Fig.~\ref{fig:noise}{\bf b}.

%----------------------------------------------------------------------
\section{Reliability of numerical simulation methods}
\label{sec:simulations}

Development of reliable and fast numerical simulations of the amplifier dynamics represents a challenging problem. 
Langevin equations~\eqref{eq:Langevin_explicit} contain three separated but equally important timescales determined by frequencies 
$\gamma\approx 5$ MHz, $\Omega_0 /2\pi \approx 2.9$ GHz, and $\Omega_{\rm J}/2\pi\approx 270.4$ GHz. This makes the problem stiff. Consequently, the numerical simulations based on usual non-stiff schemes can be extremely inefficient. They would require simultaneously a very small integration time-step, to catch the fastest oscillations and stabilize the integration of stiff differential equations, and long absolute times to describe correctly the slowest time scale. The fact that we are interested in the noise power spectrum~\eqref{eq:S_vout}, i.e., in the Fourier transform of the time-correlation function, makes the simulations even more demanding. 

%%%%%%%%%%%%%%%%%%%%%%%%%%%%%%%%%%%%%%%%%%%%
\begin{figure}[t!]
	\centering
	\includegraphics[width=0.75\textwidth]{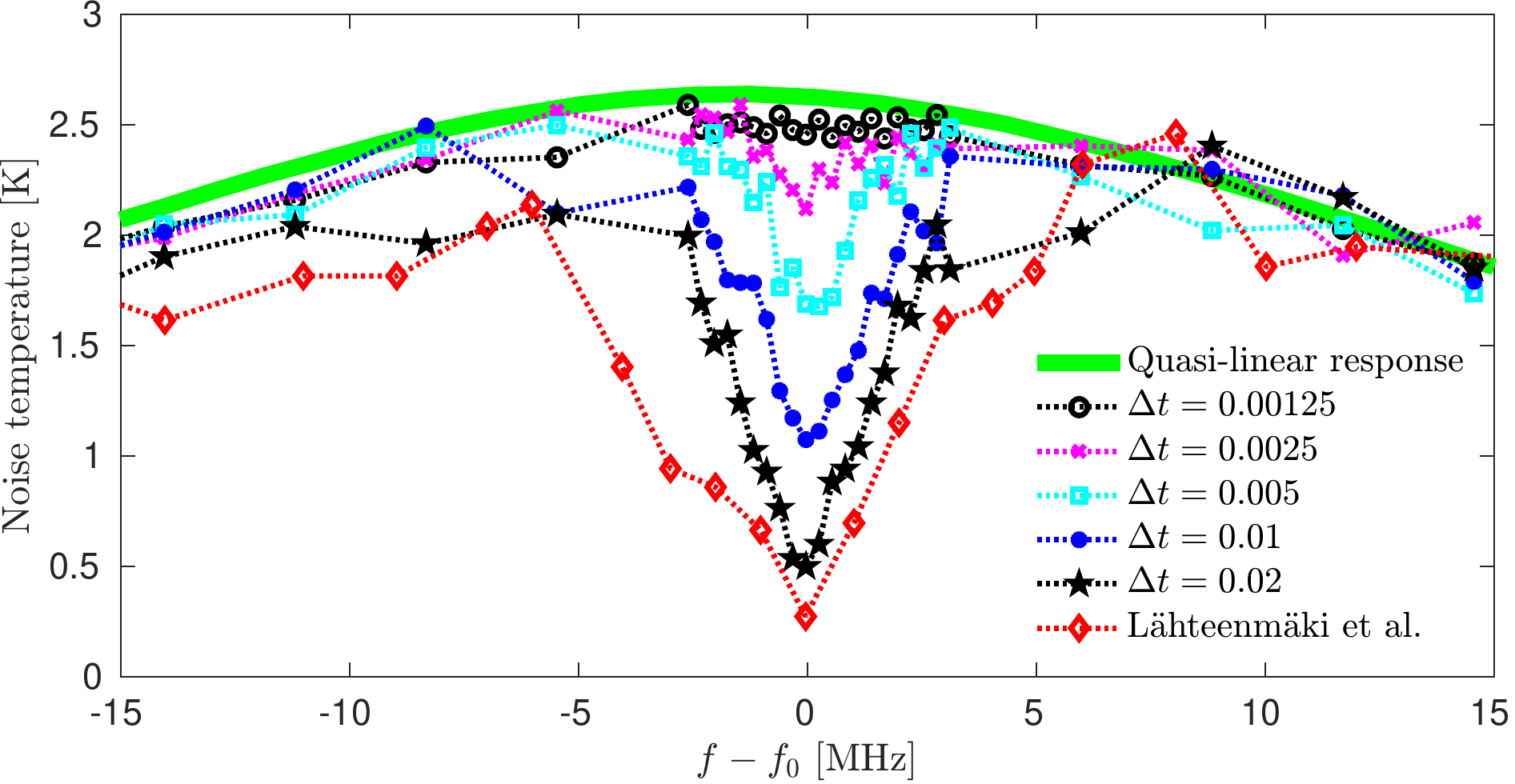} 
	\caption{Comparison of noise temperatures obtained from the quasi-linear response theory (see also Fig.~\ref{fig:noise}{\bf b}), numerical simulations using stiff integration scheme~(\ref{eq:scheme}) with different time-steps $\Delta t$, and the simulation result taken from Ref.~\cite{Lahteenmaki2012} (Lahteenmaki et al.) 
	} 
	\label{fig:simulations}
\end{figure}
%%%%%%%%%%%%%%%%%%%%%%%%%%%%%%%%%%%%%%%%%%%%

To overcome these difficulties, we have utilized a modification of the implicit Euler method \cite{Burrage/Tian:2001} 
%\begin{equation} 
%y_{n+1}=y_{n}+f(t_{n+1},y_{n+1})\Delta t+g(t_{n+1},y_{n+1})\Delta W_{n}
%\end{equation}    
\begin{equation} 
\label{eq:scheme}
{\bm x}_{n+1}= {\bm x}_{n}
+[{\bm f}({\bm x}_{n+1})+{\bm g}v_{\rm in}(t_{n+1}) ]\Delta t 
+{\bm \xi}\Delta W_{n},
\qquad n=0, 1, 2, \ldots
\end{equation}    
where $t_n=n\Delta t$, ${\bm x}_n={\bm x}(t_n)$, and $\Delta W_{n} = \int_{t_n}^{t_{n+1}}\! \dd t \delta i(t)$. The algebraic equation for the unknown ${\bm x}_{n+1}$ is solved
using the fixed-point iteration scheme. Here, we set a very strict convergence criterion with $|{\bm x}_{n+1,i+1}-{\bm x}_{n+1,i}| < 10^{-13}$ and at least ten iterations $i$ to stop the scheme. Although this requirement significantly slowed down the simulation, it was crucial for maintaining the stability of the integration scheme. 

An additional challenge is the generation of the colored semi-classical resistor noise with the power spectrum~\eqref{eq:SI}. To this end, we have adopted the method introduced in Ref.~\cite{Billah/Shinozuka:1990} according to which the stochastic process is simulated using 
\begin{equation}
\delta i(t)= \sum_{n=1}^{N}
\left[2S_i(\Delta\omega n) \Delta\omega \right]^{\frac{1}{2}}
\cos(\Delta\omega n t + 2\pi r_{n}),
\end{equation}
where $\Delta\omega=\omega_{u}/N$, $\omega_{u}$ is the upper cut-off frequency and $r_{n}$ are
random numbers uniformly distributed in the interval [0,1]. An advantage of this method is that it allows to speed up the calculation of noise by employing the fast-Fourier transform technique. The total duration of each simulation was set between $10^{6}/\Omega_0$ and $10^{7}/\Omega_0$. 
Note, that such long simulation times are necessary for obtaining a converged noise power spectrum. This is a consequence of the longest time scale governed by $\gamma$ ($\Omega_0 /\gamma \approx 3000$). In addition, the presented data were averaged over $3000$ independent trajectories. Yet, the results still show a strong dependence on the integration time-step $\Delta t$.

This is illustrated in Fig.~\ref{fig:simulations} where we plot simulated noise temperature for model parameters in Tab.~\ref{table} and five various integration time-steps in dimensionless units. In compliance with Ref.~\cite{Lahteenmaki2012}, larger integration steps lead to a significant dip in the noise temperature around the working frequency. However, this feature vanishes with decreasing $\Delta t$. For the smallest time-step $\Delta t=0.00125$, the dip disappears and simulation results are in a good agreement with the result of quasi-linear response theory. 

As discussed in Sec.~\ref{sec:quasilinear} [bellow Eq.~\eqref{eq:Dtheta_Si}], any low-frequency noise can contribute significantly to values of $D_\vartheta$ both in the experiment and simulations. Remarkably, the discretization errors in integration of Langevin equations can represent just such a noise contribution and can, therefore, significantly influence the simulated noise temperature even at relatively small integration time-steps. The resulting noise suppression effect is then qualitatively similar to the one shown in Fig.~\ref{fig:noise}{\bf b}. 

%----------------------------------------------------------------------
\section{Summary and perspectives}
\label{sec:summary}

We have analyzed stochastic nonlinear dynamics of a circuit corresponding to the microwave amplifier working close to the quantum limit. The amplifier was designed in experimental works~\cite{Lahteenmaki2012} and \cite{Lahteenmaki2014} based on a circuit involving an unshunted Josephson junction as the amplification element. We have developed the quantitative theory explaining physical origins and magnitudes of the amplifier gain and the noise power spectrum and addressed the precision and limitations of numerical schemes commonly used for integrating the underlying nonlinear Langevin equations. 

If the Josephson junction is biased by an overcritical current, its phase starts to oscillate forcing the circuit dynamics to settle down onto a stable limit cycle. In such a parameter regime, the circuit can act as a microwave amplifier. We have shown that its amplification capability is set up by amplitudes of slowly-decaying Floquet modes excited by the weak input signal and by thermal fluctuations of voltage at a resistor involved in the circuit. Decay rates of these modes are given by associated Floquet exponents and are inherently related to the stability of the limit cycle. We propose to control the stability by varying the bias current in the experiments which also provides a direct control over the amplifier gain and bandwidth. Namely, the less stable the limit cycle is (the bias current close to the critical current), the higher the amplifier gain can be. The bandwidth decreases with the approach to instability such that the maximum gain-bandwidth product remains constant. 

Thermal fluctuations at the resistor stimulate the diffusion of the limit cycle phase, which alters the principal device characteristics. Its impact on the gain curve is most pronounced close to the limit cycle instability. There, the phase fluctuations cause the decrease of the maximum gain point and increase the bandwidth compared to theoretical predictions where the phase diffusion is neglected.  

Furthermore, we have derived the expression for the noise power spectrum at the amplifier output port. The formula holds for an arbitrary colored resistor noise and can include multiple independent noise sources. 
Assuming the Callen-Welton spectrum for the parameters of the experiments in Ref.~\cite{Lahteenmaki2012}, it implies that quantum effects (reflected by the linear part of the spectrum) contribute significantly to the enhancement of the output noise spectral power. If one assumes a classical Johnson-Nyquist noise instead, the intrinsic noise power is by several orders of magnitudes smaller than the one reported in the experimental works. Again, close to instability, the phase diffusion contributes to the width of the intrinsic noise power. However, the noise temperature remains constant as broadening of the gain and the noise curves induced by the phase diffusion is the same close to the gain peak. 

As for the numerics, we have discussed precision of the explicit and implicit numerical schemes for the integration of nonlinear Langevin equations. Since the system of equations is stiff, reliable and computationally expensive implicit schemes with extremely small discretization steps have to be employed to achieve an agreement with theory. Otherwise, the discretization errors could create apparent physical effects manifesting themselves as a deep minimum in the noise temperature close to the amplifier working frequency. A plausible explanation of this observation is based on properties of the phase diffusion process. In course of integration, the random discretization errors can accumulate into an apparent low-frequency noise source that increases the phase diffusion coefficient beyond its actual value. Such an increase implies broadening of the noise spectrum and the corresponding dip in the noise temperature can occur. 

This analysis demonstrates an advantage of our analytical results when compared to the numerical integration of Langevin equations. Computations based on the derived analytical formulas are significantly faster to evaluate in practice and hence they can be used to optimize performance of a particular circuit. Also, the theoretical approach is more reliable as it does not introduce the aforementioned unwanted artifacts close to the amplifier optimal operation point. Of course in practice, it is desirable to employ both complementary methods (the theory and simulations) to verify performance characteristics for at least a single set of experimentally relevant parameters. 

Finally, even though our analysis is inspired by a particular circuit, the theoretical approach and derived results are general enough to be used to design a wide range of high-gain and low-noise amplifiers based on nonlinear circuit elements and operating near a stable limit cycle. In this respect, the theory provides physically motivated instructions how to optimize performance of such circuits in presence of arbitrary internal and external weak colored noises. In addition to these applications in physics, our method can be straightforwardly applied in other areas of nonlinear science like mathematical biology and ecology where colored noises and harmonic signals perturb dynamics of a stable limit cycle. There, our methodology can be utilized to provide a fast and reliable algorithm for computation of system's power spectra and response curves. 

%----------------------------------------------------------------------
\section*{Acknowledgements}
\noindent 
This work was supported by the Czech Science Foundation projects No.~20-24748J~(AR) and No.~19-13525S~(TN),  by the grant INTER-COST LTC19045~(M{\v Z}), by the COST Action NANOCOHYBRI (CA16218) (TN), and by the National Science Centre (NCN, Poland) via the grant No.\ UMO-2017/27/B/ST3/01911~(TN). 

%----------------------------------------------------------------------
% Reference 
%\bibliography{amplifier_references}

\end{document}